\documentclass[aps,prl,preprint]{revtex4-1}

\usepackage[utf8]{inputenc} 
\usepackage{amsmath,amssymb}
\usepackage{graphicx}
\usepackage{booktabs}
\usepackage{multirow}
\usepackage{makecell}
\usepackage{tabularx}
\usepackage[version=4]{mhchem}
\usepackage{xcolor} 
\usepackage{natbib} 
\usepackage{adjustbox}
\usepackage{float}
\usepackage{array}[=2016-10-06]
\usepackage{tabularx}
\usepackage{braket}
\usepackage{hyperref}
\usepackage{placeins}

\newcommand{\scite}[1]{\textsuperscript{\citep{#1}}}
\newcolumntype{C}[1]{>{\centering\arraybackslash}p{#1}}

\begin{document}

\title{Complex-Energy Second-Order Approximate\\ Coupled-Cluster Methods for Electronic Resonances}

\author{Cansu Utku}
\affiliation{Department of Chemistry, KU Leuven, Belgium}

\author{Garrette Pauley Paran}
\affiliation{Department of Chemistry, KU Leuven, Belgium}

\author{Thomas-C. Jagau}
\affiliation{Department of Chemistry, KU Leuven, Belgium}

\begin{abstract}
Electronic resonances are metastable states with finite lifetimes, encountered in processes such as photodetachment, electron transmission, and Auger decay. Resonances appear in Hermitian quantum mechanics as increased density of states in the continuum rather than as discrete energy levels. To describe resonances accurately, including their coupling to the continuum, methods based on non-Hermitian quantum mechanics can be used, which yield complex energies. In this work, we combine the complex absorbing potential (CAP) and complex basis functions (CBF) techniques with the RI-CC2 method. The second-order coupled cluster method (CC2) offers a good balance between accuracy and computational cost by approximating equation-of-motion coupled-cluster singles and doubles (EOM-CCSD) theory, making it suitable for studying of electronic resonances in larger molecules. The resolution-of-the-identity (RI) approximation further reduces computational demands without significant loss in accuracy. We investigate the numerical performance of the new complex-energy RI-CC2 methods focusing on temporary anions. Negative electron affinities and decay widths can be computed using the electron-attachment (EA) variant of RI-CC2. For N$_2$, C$_2$H$_4$, CH$_2$O, and HCOOH, EA-CC2 yields affinities about 0.1–0.2 eV smaller than EOM-EA-CCSD, while deviations reach 0.5 eV for larger anions such as uracil, naphthalene, cyanonaphthalene, and pyrene. As a result of these trends, EA-CC2 is in better agreement with experiment for the negative electron affinities than EOM-EA-CCSD for all studied anions. The corresponding resonance widths from EA-CC2 calculations are about 0.05–0.25 eV smaller compared to EOM-EA-CCSD. Semi-empirical spin-scaling increases electron affinities by 0.3-0.5 eV and broadens resonance widths, improving the agreement with EOM-EA-CCSD but worsening the agreement with experiment.

\end{abstract}

\date{\today}
\maketitle


\section{Introduction} \label{sec:in}

The characterization of metastable electronic states plays an important role in understanding 
processes such as dissociative electron attachment\citep{Simons2008,fabrikant2017recent,Simons2008}, 
autoionization, Auger-Meitner decay\citep{meitner1922beta,Auger1923}, and intermolecular Coulombic 
decay\citep{cederbaum1997giant}.These processes mediate key steps in chemical reactions including, 
for example, energy transfer,\citep{costentin2006electron,studer2014electron} plasmonic catalysis,\citep{christopher2011visible,mukherjee2013hot,christopher2012singular} and radiation damage in 
materials and biological systems.\citep{boudaiffa2000resonant,alizadeh2015biomolecular} Metastable states are also essential for interpreting 
Auger,\citep{agarwal2013x} photoionization,\citep{simons2000detachment} photodetachment,\citep{sanov2014laboratory}, electron scattering,\citep{mcvoy1968virtual,taylor2012scattering} 
and electron-energy loss spectra.\citep{allan2016recent} Their study and understanding are thus essential for 
advancing fields ranging from quantum chemistry and spectroscopy to plasma physics, astrochemistry, 
and biomedical science. 

Metastable states lie in the continuum above the detachment or ionization threshold and cannot be 
described within the conventional bound-state framework. Electronic resonances have been studied 
using a variety of theoretical approaches that include both extensions of bound-state techniques and 
explicit scattering methods. Stabilization methods identify resonances by tracking the variation of 
eigenvalues upon introducing external charges or scaling the diffuseness of basis functions.\citep{
Hazi1970,nestmann1985calculation} The Feshbach-Fano formalism partitions the Hilbert space into 
bound and continuum subspaces and treats the coupling perturbatively.\citep{Feshbach1962,Fano1961} 
Scattering-based techniques such as the R-matrix\citep{Tennyson2010} method solve the electron-molecule 
scattering problem directly, yielding resonance positions and widths from the scattering matrix.\citep{Taylor1972} 

A further approach for the theoretical treatment of resonances involves non-Hermitian extensions of 
quantum chemistry methods.\citep{moiseyev2011non,Jagau2022} Complex absorbing potentials\citep{
jolicard1985optical,riss1993calculation} (CAPs) and complex basis functions (CBFs)\citep{
mccurdy1978extension,moiseyev1979autoionizing,white2015complex} provide a route to compute 
resonance energies and lifetimes within the framework of bound-state electronic structure methods. 

These approaches yield complex energy and have been implemented for a variety of electronic-structure 
methods. CAPs have been employed with density functional theory (DFT),\citep{zhou2012calculating} multireference 
configuration interaction (MRCI)\citep{sommerfeld1998temporary,sommerfeld2001efficient}, Fock space 
multireference coupled-cluster,\citep{sajeev2005general} symmetry-adapted cluster configuration interaction 
(SAC-CI)\citep{ehara2012cap}, algebraic diagrammatic construction [ADC($n$)]\citep{santra2002complex,
feuerbacher2003complex,belogolova2021complex,dempwolff2021cap}, second-order multiconfigurational 
quasidegenerate perturbation theory (XMCQDPT2),\citep{kunitsa2017cap} complete-active-space second-order 
perturbation theory (CASPT2),\citep{phung2020combination} and equation-of-motion coupled-cluster with singles and 
doubles (EOM-CCSD).\citep{ghosh2012equation,jagau2014fresh,zuev2014complex,gayvert2022application,
camps2025complex} The CAP-EOM-CCSD method has been extended beyond energies and lifetimes 
to transition properties\citep{jagau2016characterizing} and analytic gradients.\citep{benda2017communication,benda2018understanding,
mondal2025analytic}

We note that there are different variants of CAP-SAC-CI, CAP-ADC, and CAP-EOM-CCSD available, 
in which the potential is added to the Hamiltonian at different steps of the calculation. For example, in 
EOM-CCSD, one can either include the CAP in the Hartree-Fock (HF) calculation \citep{zuev2014complex}, 
or add it only at the EOM-CCSD step\citep{ghosh2012equation}, or construct the CAP Hamiltonian in a 
basis of real-valued EOM-CCSD pseudocontinuum states.\citep{gayvert2022application} By means of 
the OpenCAP software,\citep{gayvertopencap} the latter approach has been extended to arbitrary 
electronic-structure methods that yield mutually orthogonal pseudocontinuum states.

The CBF method relies on complex scaling\citep{aguilar1971class,balslev1971spectral,bravaya2013complex} 
(CS) and has been combined with HF theory,\citep{rescigno1980application,mccurdy1980applicability,white2015complex} MRCI,\citep{
honigmann1999complex,honigmann2006complex,honigmann2010use} resolution-of-the-identity second-order 
M{\o}ller-Plesset perturbation theory (RI-MP2), \citep{hernandez2019resolution,hernandez2020resolution} 
and EOM-CCSD.\citep{white2017second,camps2025complex} 

In this work, we present an implementation of the electron attachment (EA) variant of the resolution-of-the-identity 
second-order approximate coupled-cluster (RI-CC2) method\citep{christiansen1995second,hattig2000cc2,hattig2006beyond} combined with both CAPs and CBFs. In analogy to CC2 for bound states, which is suited well 
for electronically excited states in large systems, our implementation provides an appealing alternative to 
EOM-CCSD for resonances in larger systems. The computational cost scales as $\mathcal{O}(N^5)$ with 
system size $N$ as compared to $\mathcal{O}(N^6)$ for EOM-CCSD, balancing accuracy and efficiency. 
The memory requirements are reduced from $\mathcal{O}(N^4)$ to $\mathcal{O}(N^3)$ by decomposing 
the four-index two-electron integrals by means of the RI approximation into three-index quantities.\citep{
whitten1973coulombic,feyereisen1993use,weigend1998ri} We note that the latter can also be achieved 
by means of Cholesky decomposition (CD)\citep{beebe1977simplifications,koch2003reduced,
aquilante2011cholesky}, but this is beyond the present work. 

Our RI-CC2 implementation is based on the algorithm presented in Ref.~\citep{hattig2000cc2}. This algorithm 
is based on an effective Jacobian that only has the dimensions of the CI singles eigenvalue equations 
because the double excitation amplitudes are treated implicitly. As a result, the matrix elements of the 
Jacobian depend on the excitation energy, i.e., the eigenvalue that is sought. Therefore, the Davidson 
algorithm, commonly applied to solve the EOM-CCSD equations, cannot be applied to RI-CC2 without 
modifications. To address this, we implemented a modified algorithm that is also based on Ref.~\citep{hattig2000cc2}. 

Extending the original formulation for electron-number and spin conserving excitations,\citep{
christiansen1995second,hattig2000cc2} further RI-CC2 methods have been implemented for spin-flipping 
(SF) excitations,\citep{paran2022spin} ionization (IP),\citep{walz2016hierarchy} and electron attachment.\citep{
ma2020approximate} The numerical performance of the IP and EA variants is not satisfactory, but it can be 
improved significantly using spin scaling factors\citep{grimme2012spin} resulting in the spin-component-scaled 
(SCS)\citep{grimme2003improved} and spin-opposite-scaled (SOS)\citep{jung2004scaled} RI-CC2 methods.\citep{
hellweg2008benchmarking} For example, in the case of valence anions, the error of RI-EA-CC2 relative to 
EOM-EA-CCSD is 0.47 eV, but decreases to 0.07 eV with SCS and 0.14 eV with SOS.\citep{paran2024performance} 
Very similar results were found for the spin-scaled ADC(2) methods.\citep{shaalan2022accurate} This suggests 
that complex-energy RI-EA-CC2 may also benefit from spin scaling and, therefore, we also implemented the 
SCS and SOS variants. 

To assess the performance of complex-energy RI-CC2 methods for resonance states, we apply our new 
methods to several molecular temporary anions. These include the small anions N$_2^-$, C$_2$H$_4^-$, 
CH$_2$O$^-$, HCOOH$^-$, for which EOM-EA-CCSD are feasible, as well as anions of uracil, naphthalene, 
cyanonaphthalene, and pyrene as examples of larger molecules. The anion of uracil is of interest in 
the context of radiation-induced damage to RNA. The anions of the polycyclic aromatic hydrocarbons (PAHs) 
are astrochemically relevant,\citep{tielens2008interstellar} because PAHs are widespread in the interstellar 
medium and believed to contribute to diffuse interstellar bands.

The remainder of the article is structured as follows: In Section \ref{sec:th}, we discuss the theory of 
complex-energy techniques and the RI-CC2 electronic-structure method. Thereafter, we present the 
details of our implementation in Section \ref{sec:imp}. The computational details and numerical results 
of the application of RI-CC2 to temporary anions are presented in Sections \ref{sec:cd} and \ref{sec:res}, 
respectively. Our conclusions about the new complex-energy RI-CC2 methods are given in Section 
\ref{sec:con}.


\section{Theory} \label{sec:th}
\subsection{Complex-energy methods} \label{sec:ce}

The accurate description of electronic states embedded in the detachment continuum requires theoretical 
methods that go beyond a bound-state treatment. Complex-energy methods offer a systematic framework 
to characterize metastable states in terms of their energy and lifetime. These methods rely on a non-Hermitian 
formulation of the Schr\"odinger equation, achieved either by adding a CAP\citep{jolicard1985optical,
riss1993calculation} or by complex scaling the coordinates of the Hamiltonian\citep{aguilar1971class,
balslev1971spectral} or the exponents of Gaussian basis functions.\citep{mccurdy1978extension,
moiseyev1979autoionizing} The resulting complex eigenvalues of the Schr\"odinger equation are called 
Siegert energies and can be expressed as:\citep{siegert1939derivation}
\begin{equation} \label{eq:siegert}
E = E_R - i \, \Gamma/2
\end{equation}
where the real part $E_R$ corresponds to the energy of the resonance and the imaginary part $\Gamma/2$ 
is connected to the lifetime $\tau$ of the metastable state via $\Gamma = 1/\tau$. In the context of temporary 
anion states, $E_R$ is also referred to as resonance position and the difference between $E_R$ and the 
energy of the parent neutral molecule yields the electron affinity, which is negative. 

\subsection{Complex absorbing potentials} \label{sec:cap}

The CAP Hamiltonian has the form:
\begin{equation} \label{eq:capH}
H_{\eta} = H - i \, \eta \, W,
\end{equation}
where $\eta$ is the CAP strength and $W$ defines the shape of the absorbing potential. In this work, we 
employ a quadratic box potential
\begin{equation} \label{eq:cap}
W = \sum_\alpha W_\alpha ~, \quad W_{\alpha}(r_\alpha) =
\begin{cases}
0 & \text{if} \, |r_{\alpha} - o_\alpha| \leq r^0_{\alpha}~, \\
(|r_{\alpha} - o_\alpha| - r^0_{\alpha})^2 & \text{if} \, |r_{\alpha} - o_\alpha| > r^0_{\alpha}~. 
\end{cases}
\end{equation}
Here, $\alpha = x, y, z$ refers to the Cartesian coordinates, $o_\alpha = (o_x,o_y,o_z)$ is the origin of 
the CAP, and $r^0_{\alpha}$ is the size of the box. We also compare to results computed with a smooth 
Voronoi potential, where each atom is enclosed within a cutoff sphere.\citep{sommerfeld2015complex,
gayvert2022application} The smooth Voronoi CAP has the form
\begin{equation} \label{eq:cap2}
W(r) = 
\begin{cases}
\left( r_{\text{av}}(r) - r^0 \right)^2 & \text{if } r_{\text{av}} > r^0~, \\ 0 & \text{else}~,
\end{cases} \qquad 
r_{\text{av}}(r) = \sqrt{\frac{\sum_A w_A(r) \, r_A^2(r)}{\sum_A w_A(r)}}
\end{equation}
and depends on a single parameter $r_0$. The index $A$ refers to the nuclei, $w_A$ are the weighting 
factors, and $r_\text{av}$ is the weighted average of the electron-nucleus distances.

The CAP absorbs the diverging tail of the resonance wave function, forcing it into a square-integrable form. 
The eigenvalues of the CAP Hamiltonian depend on $\eta$ as well as $r^0_\alpha$. In this work, we derive 
the onset parameters $r^0_\alpha$ from the extent of the wave function of the neutral ground state\citep{
jagau2014fresh} and optimize $\eta$ according to the criterion $\min |\eta \, (dE / d\eta) |$.\citep{riss1993calculation}
This necessitates series of calculations with different values of $\eta$. This parameter has no strict upper 
bound and optimal values can range from $10^{-4}$ a.u. to $10^{-1}$ a.u. In the CAP calculations presented 
in this work, the trajectory is sampled over 30--40 $\eta$ values, ranging from 0.00025 to 0.01 a.u. with 
increments of 0.00025 a.u. An alternative approach, in which the CAP strength $\eta$ is fixed and the 
onset parameters $r^0_\alpha$ are optimized, has been introduced recently,\citep{gyamfi2024new} and entails 
a similar number of calculations. 

The need to compute trajectories and the resulting computational overhead motivate the projected CAP 
technique.\citep{sommerfeld2001efficient} In this approach, the Schr\"odinger equation is first solved without 
CAP for a number of pseudocontinuum states and the CAP Hamiltonian from Eq.~\eqref{eq:capH} is then 
constructed in the basis of these states. This significantly reduces the cost of generating the $\eta$ trajectory 
as only one electronic-structure calculation needs to be carried out. However, the resulting Siegert energies 
depend on the size of the subspace. The required number of states is system dependent and cannot be 
predicted \textit{a priori}: For example, for projected CAP-EOM-EA-CCSD, 4 pseudocontinuum states 
were sufficient for N$_2^-$, whereas anionic states of uracil required as many as 55 states.\citep{gayvert2022projected}

Lastly, we note that Siegert energies computed with CAP methods can often be improved somewhat 
by means of perturbation theory, i.e., by subtracting the expectation value of $ -i \, \eta \, W$ from the 
corresponding eigenvalue of $H_\eta$.\citep{riss1993calculation,jagau2014fresh} In Section \ref{sec:res}, these 
results are referred to as ``first-order''.


\subsection{Complex basis functions} \label{sec:cbf}

The method of complex basis functions (CBFs)\citep{mccurdy1978extension,white2015complex} extends 
CS\citep{aguilar1971class,balslev1971spectral} to molecular resonances. This is achieved by using 
a Gaussian basis set in which the exponents $\alpha$ of some functions $\chi_A$ are complex-scaled 
yielding $\alpha e^{-2i\theta}$ with $\theta$ as scaling angle:
\begin{equation} \label{eq:cbf}
\chi_A (r, \theta) = N(\theta) \, (x - x_A)^{k} \, (y - y_A)^{l} \, (z - z_A)^{m} \, \exp 
\big[ -\alpha e^{-2 i \theta} \, (r - R_A )^2 \big]
\end{equation}
$N(\theta)$ is a normalization constant and $R_A$ the position of nucleus $A$. Whereas the CAP 
method has been used primarily for temporary anions, the CBF method has been applied to various 
types of resonances. This includes molecules in static electric fields\citep{jagau2018coupled,
thompson2019schwarz,hernandez2019resolution,hernandez2020resolution}, Rydberg states\citep{
creutzberg2023computing,camps2025complex}, Auger decay\citep{matz2022molecular,
matz2023channel,jayadev2023auger,matz2025complex}, and intermolecular Coulombic decay\citep{
parravicini2023interatomic} in addition to temporary anions.\citep{white2015complex,white2015restricted,
white2017second} The exponents of the complex-scaled functions are related to the kinetic energy 
of the emitted electrons,\citep{matz2022molecular} meaning that diffuse functions need to be scaled 
for low-lying temporary anions and steeper functions for Auger decay.
 
The complex scaling angle $\theta$ is restricted to $0 < \theta < \pi/4$,\citep{moiseyev2011non} which is a practical 
advantage over the CAP method where no upper bound exists for the CAP strength $\eta$. Nonetheless, 
an optimal scaling angle $\theta_{\mathrm{opt}}$ must be determined by minimizing the variation of the 
resonance energy with respect to changes in $\theta$, following the condition $\min | dE/d\theta |$.\citep{moiseyev1978resonance} 
Identifying $\theta_{\mathrm{opt}}$ requires multiple calculations with different $\theta$ values, For the 
temporary anions studied in the present work, $\theta_{\mathrm{opt}}$ was determined by 
scanning the range from $0^\circ$ to $45^\circ$ in increments of $4^\circ$ to identify an approximate 
value, followed by a finer search in the vicinity of this estimate. This strategy reduces the number of 
calculations required to determine $\theta_{\mathrm{opt}}$ compared to CAP calculations. Often, only 
10--12 calculations are needed. Also, the minima in $| dE/d\theta |$ are typically more pronounced 
than those in $| \eta \, dE/d\eta |$ for CAP calculations. 


\subsection{The RI-CC2 reference state} \label{sec:ricc2}

CC2 is a second-order approximate coupled-cluster method in which the single amplitude equations 
are identical to those from CCSD and the double amplitude equations are truncated to first order in 
perturbation theory.\citep{christiansen1995second} A general CC wave function is expressed using 
an exponential ansatz as 
\begin{equation} \label{eq:ccwf}
|\Psi_\text{CC} \rangle = e^T | \Phi_0 \rangle
\end{equation}
with $|\Phi_0 \rangle$ as the reference determinant and the cluster operator defined as $T = T_1 + T_2 
+ \cdots + T_n$ for $n$ electrons. For CCSD and CC2 alike, $T$ is truncated to the singles and 
doubles cluster operators, which are given by
\begin{equation} \label{eq:t}
T_1= \sum_{ai} t_i^a \, a^\dagger i ~, \quad
T_2  = \frac{1}{4} \sum_{abij} t_{ij}^{ab} \, a^\dagger i b^\dagger j 
\end{equation}
Here, $i, j, \dots$ denote occupied molecular orbitals, while $a, b, \dots$ refer to virtual orbitals. $p, q, 
\dots$ will be used to denote generic orbitals. In second quantization notation,\citep{jorgensen2012second} 
$a^\dagger $ creates a particle in spin orbital $\varphi_a$ and $i$ removes a particle from spin orbital 
$\varphi_i$ and thus creates a hole. 

The amplitudes $t_i^a$ and $t_{ij}^{ab}$ are determined from the CC equations, which are derived by  
inserting Eq. \eqref{eq:ccwf} into the Schr\"odinger equation and applying $e^{-T}$ from the left, followed 
by projection on the singly and doubly excited determinants,
\begin{align} \label{eq:ccs}
\Omega_i^a &= \langle \Phi_i^a | e^{-T} H e^{T} | \Phi_0 \rangle = 0  \quad \forall ~ \langle \Phi_i^a | ~, \\
\Omega_{ij}^{ab} &= \langle \Phi_{ij}^{ab} | e^{-T} H e^{T} | \Phi_0 \rangle = 0  \quad \forall ~ \langle \Phi_{ij}^{ab} | ~, 
\label{eq:ccd}
\end{align}
where $e^{-T} H e^{T}$ is the similarity-transformed Hamiltonian. The CC energy is obtained from 
projection onto the reference determinant, 
\begin{equation} \label{eq:cce}
E_\text{CC} = \langle \Phi_0 | e^{-T} H e^T | \Phi_0 \rangle = E_0 + \sum_{ai} F_a^i \, t_i^a 
+ \frac{1}{2} \sum_{ijab} \langle ij || ab \rangle \, t_i^a t_j^b 
+ \frac{1}{4} \sum_{ijab} \langle ij || ab \rangle \, t_{ij}^{ab}~,
\end{equation}
where $F_a^i$ is an element of the Fock matrix and $\langle ij || ab \rangle$ a two-electron repulsion 
integral. 

In CC2 theory, Eq. \eqref{eq:ccd} is simplified based on a perturbative analysis. The Hamiltonian
\begin{equation} \label{eq:h}
H =  E_0 + F + W = E_0 + \sum_{pq} F_p^q \, \{ p^\dagger q \} 
+ \frac{1}{4} \sum_{pqrs} \langle pq || rs \rangle \, \{ p^\dagger q^\dagger sr \}
\end{equation}
is partitioned into a zeroth-order term $E_0 + F$ and a first-order term $W$. $T_1$ and $T_2$ are 
assigned to zeroth order and first order, respectively. Keeping only terms that are at most first order 
in Eq. \eqref{eq:ccd} yields the CC2 double amplitude equations 
\begin{equation} \label{eq:doubles}
\Omega_{ij}^{ab} = \langle \Phi_{ij}^{ab} | \tilde{H} + [F, T_2] | \Phi_0 \rangle = 0 \quad \forall ~ \langle \Phi_{ij}^{ab} | \\
\end{equation}
where $\tilde{H} = e^{-T_1} H e^{T_1}$ is a Hamiltonian similarity-transformed by $T_1$ and has the 
same particle rank as $H$. An explicit expression in terms of spin orbitals is
\begin{align} \nonumber
0 &= \langle ab || ij \rangle + P(ab) \sum_c F_c^b t_{ij}^{ac} - P(ab) \sum_{kc} F_c^k \, t_k^b \, t_{ij}^{ac}  
- P(ij) \sum_k F_j^k t_{ik}^{ab} - P(ij) \sum_{kc} F_c^k \, t_j^c \, t_{ik}^{ab} \\ \nonumber
&- P(ab) \sum_k \langle kb || ij \rangle \, t_k^a + P(ij) \sum_c \langle ab || cj \rangle \, t_i^c 
+ \sum_{cd} \langle ab || cd \rangle \, t_i^c \, t_j^d + \sum_{kl} \langle kl || ij \rangle \, t_k^a \, t_l^b \\ \nonumber
&- P(ij|ab) \sum_{kc} \langle kb || cj \rangle \, t_i^c \, t_k^a + P(ab) \sum_{kcd} \langle kb || cd \rangle \, 
t_i^c \,  t_k^a \, t_j^d + P(ij) \sum_{klc} \langle kl || cj \rangle \, t_i^c \, t_k^a \, t_l^b \\ 
&+ \sum_{klcd} \langle kl || cd \rangle \, t_i^c \, t_j^d \, t_k^a \, t_l^b \label{eq:cc2d} \\ 
&= \langle \hat{a} \hat{b} || \hat{i} \hat{j} \rangle + P(ab) \sum_c \tilde{F}_c^b \, t_{ij}^{ac} 
- P(ij) \sum_k \tilde{F}_j^k \, t_{ik}^{ab} \quad \forall \; i>j, \; a>b \label{eq:cc2d1}
\end{align}
Here, $P(ab)$, $P(ij)$, and $P(ij|ab)$ are antisymmetric permutation operators and $\langle 
\hat{a} \hat{b} || \hat{i} \hat{j} \rangle$ denotes the $T_1$-transformed electron-repulsion integrals that 
arise as matrix elements of $\tilde{W} = e^{-T_1} W e^{T_1}$. Note that for a generic electron-repulsion 
integral $\langle \hat{p}\hat{q} || \hat{r}\hat{s} \rangle$, the $T_1$-transformation affects $p$ and $q$ 
only if they are virtual orbitals and $r$ and $s$ only if they are occupied orbitals because $T_1$ is a 
pure excitation operator.

$\tilde{F}$ is the $T_1$-transformed Fock matrix that is constructed from the $T_1$-transformed one-electron 
Hamiltonian $\tilde{h} = e^{-T_1} h \, e^{T_1}$ and $\tilde{W}$. If $| \Phi_0 \rangle$ is a canonical 
HF wave function, Eq. \eqref{eq:cc2d1} can be rearranged to obtain an explicit expression for 
the double amplitudes as a function of the single amplitudes as 
\begin{equation} \label{eq:cc2d2}
t_{ij}^{ab} = \frac{\langle \hat{a} \hat{b} || \hat{i} \hat{j} \rangle}{\varepsilon_i + \varepsilon_j 
- \varepsilon_a - \varepsilon_b}
\end{equation}
The significance of Eq. \eqref{eq:cc2d2} is that it enables an implementation in which the double 
amplitudes $t_{ij}^{ab}$ need not be stored. 

The CC2 single amplitude equations are identical to those from CCSD and can be written as 
\begin{equation} \label{eq:singles}
\Omega_i^a = \langle \Phi_i^a | \tilde{H} + [\tilde{H}, T_2] | \Phi_0 \rangle = 0 \quad \forall ~ \langle \Phi_i^a | ~,
\end{equation}
with the explicit expression 
\begin{align} \nonumber 
0 &= F_i^a + \sum_{kc} F^k_c t_{ik}^{ac} + \frac{1}{2} \sum_{kcd} \langle ak || cd \rangle \, t_{ik}^{cd}
- \frac{1}{2} \sum_{klc} \langle kl || ic \rangle\, t_{kl}^{ac} + \sum_c F_c^a \, t_i^c - \sum_k F_i^k \, t_k^a \\ \nonumber
&+ \sum_{kc} \langle ak || ic \rangle\, t_k^c -  \frac{1}{2} \sum_{klcd} \langle lk || cd \rangle \, t_l^a \, t_{ik}^{cd} 
- \frac{1}{2} \sum_{klcd} \langle kl || dc \rangle \,  t_i^d \, t_{kl}^{ac} + \sum_{klcd} \langle kl || cd \rangle\, 
t_k^c \, t_{li}^{da} \\ \label{eq:cc2s1}
&- \sum_{kc} F_c^k \, t_i^c \, t_k^a + \sum_{kcd} \langle ak || cd \rangle \, t_i^c \, t_k^d - \sum_{klc} 
\langle kl || ic \rangle \, t_k^a \, t_l^c - \sum_{klcd} \langle kl || cd \rangle \, t_i^c t_k^a t_l^d \\ \label{eq:cc2s2}
&= \tilde{F}_i^a + \sum_{kc} \tilde{F}_c^k t_{ik}^{ac} + \frac{1}{2} \sum_{kcd} \langle \hat{a}k || cd \rangle \, t_{ik}^{cd} 
- \frac{1}{2} \sum_{klc} \langle kl || \hat{i}c \rangle\, t_{kl}^{ac} \quad \forall \; i, a
\end{align}

To avoid the computation of the four-index electron-repulsion integrals, the RI approximation can be applied 
to Eqs. \eqref{eq:cc2d2}, \eqref{eq:cc2s2}, and \eqref{eq:cce}.\citep{vahtras1993integral,whitten1973coulombic,
baerends1973self,baerends1973self} A generic four-index integral, written as $( pq | rs )$ using Mulliken notation, 
is decomposed as
\begin{equation} \label{eq:ri}
(pq | rs ) = \sum_P B_{pq}^P \, B_{rs}^P = \sum_{PQ} (pq | P) \, J^{-1}_{PQ} \, (Q | rs)~, \quad 
B^{P}_{pq} = \sum_{Q} (pq | Q) \, J_{QP}^{-1/2}
\end{equation}
Here, $P$ and $Q$ denote auxiliary basis functions and $(pq|P)$ and $J_{PQ}$ are, respectively, three-center 
and two-center electron-repulsion integrals.

\subsection{Computation of excitation energies with EOM-EE-CC2 theory} \label{sec:eomcc2}

From the time evolution of the CC2 wave function, excited-state energies and state and transition 
properties can be computed within the framework of CC linear-response theory.\citep{
monkhorst1977calculation,dalgaard1983some,koch1990coupled,koch1995excitation,
christiansen1998response} However, similar to other CC models, the same working equations 
for the excitation energies can also be obtained in the framework of EOM-CC theory. Here, an 
excitation operator $R$ acts on the CC reference wave function to generate the target state 
$| \Psi_k \rangle$
\begin{equation} \label{eq:eomcc}
| \Psi_k \rangle = R_k |\Psi_\text{CC} \rangle = R_k \, e^T | \Phi_0 \rangle~.
\end{equation}
In EOM-CC2 theory, the excitation operator for excited states has the same form as in EOM-CCSD theory,
\begin{equation} \label{eq:eomee}
R^\text{EE} = R_0 + R_1 + R_2 = r_0 + \sum_{ai} r_i^a \, a^\dagger i + \frac{1}{4} \sum_{abij} 
r_{ij}^{ab} \, a^\dagger i b^\dagger j ~.
\end{equation}

The energies $\omega_k$ of the target states can be obtained by solving the eigenvalue equation 
\begin{equation} \label{eq:Eeom}
\mathbf{A} \, r_k = \omega_k r_k  \quad \Leftrightarrow \quad \left( \begin{matrix} \mathbf{A}_\text{SS} & 
\mathbf{A}_\text{SD} \\ \mathbf{A}_\text{DS} & \mathbf{A}_\text{DD} \end{matrix} \right)
\left( \begin{matrix} r_\text{S} \\ r_\text{D} \end{matrix} \right) = \omega \left( \begin{matrix} r_\text{S} \\ 
r_\text{D} \end{matrix} \right)
\end{equation}
where $r_k$ denotes the right eigenvector corresponding to state $| \Psi_k \rangle$ and \textbf{A} 
is the CC2 Jacobian. The elements of the latter matrix are obtained by differentiating the CC2 amplitude 
equations (Eqs. \eqref{eq:doubles} and \eqref{eq:singles}) with respect to $t_i^a$ and $t_{ij}^{ab}$. 
This yields 
\begin{equation} \label{eq:jac}
\mathbf{A} = \left( \begin{matrix} d\Omega_i^a / dt_k^c & d\Omega_i^a / dt_{kl}^{cd} \\ 
d\Omega_{ij}^{ab} / dt_k^c & d\Omega_{ij}^{ab} / dt_{kl}^{cd}
\end{matrix} \right) = \left(\begin{matrix}
\langle \Phi_i^a | \left[ \tilde{H} + [\tilde{H}, T_2], \{ c^\dagger k \} \right] | \Phi_0 \rangle & 
\langle \Phi_i^a | [\tilde{H}, \{ c^\dagger k d^\dagger l \}] | \Phi_0 \rangle \\
\langle \Phi_{ij}^{ab} | [\tilde{H}, \{ c^\dagger k \}] | \Phi_0 \rangle & 
\langle \Phi_{ij}^{ab} | [F, \{ c^\dagger k d^\dagger l \}] | \Phi_0 \rangle
\end{matrix} \right)
\end{equation}

Similar to Eq. \eqref{eq:cc2d2}, the second row of Eq. \eqref{eq:Eeom} can be rearranged to express 
the double amplitudes $r_{ij}^{ab}$ in terms of the single amplitudes $r_i^a$, because $\mathbf{A}_\text{DD}$ 
is diagonal with the diagonal elements given by $\Delta_{ij}^{ab} = \varepsilon_a + \varepsilon_b 
- \varepsilon_i - \varepsilon_j$. This results in 
\begin{equation} \label{eq:r2}
r_{ij}^{ab} = \frac{\sum_{kc} \langle \Phi_{ij}^{ab} | [ \tilde{H}, \{ c^{\dagger} k \} ] | \Phi_0 \rangle \, r_k^c}{\omega  
- \Delta_{ij}^{ab}} = \frac{\langle \overline{\hat{a}\hat{b} || \hat{i}\hat{j}} \rangle}{\omega + \varepsilon_i + \varepsilon_j
- \varepsilon_a - \varepsilon_b}
\end{equation}
where an additional set of modified electron-repulsion integrals has been introduced as 
\begin{equation}
\langle \overline{\hat{a}\hat{b} || \hat{i}\hat{j}} \rangle = \sum_{kc} r_k^c \; d\Omega_{ij}^{ab}/dt_k^c 
= \! \sum_c \! \left[ r_i^c \langle \hat{a}\hat{b} || c \hat{j} \rangle + r_j^c \langle \hat{a}\hat{b} || \hat{i} c \rangle \right] 
\! - \sum_k \! \left[ r_k^a \langle k\hat{b} || \hat{i}\hat{j} \rangle + r_k^b \langle \hat{a}k || \hat{i}\hat{j} \rangle \right] \, .
\end{equation}

Substituting Eq. \eqref{eq:r2} into the first row of Eq. \eqref{eq:Eeom} yields a pseudoeigenvalue 
equation in the space of single excitations that reads
\begin{equation} \label{eq:Aeff1}
\mathbf{A}^\text{eff}(\omega) \, r_\text{S} = \omega \, r_\text{S}
\end{equation}
with the elements of $\mathbf{A}^\text{eff}(\omega)$ given by 
\begin{align} \nonumber
\sigma_i^a &= (\mathbf{A}^\text{eff} \boldsymbol{\cdot} r)_i^a = \sum_{kc} \langle \Phi_i^a | \big[ \tilde{H} 
+ [\tilde{H}, T_2], \{ c^\dagger k \} \big] | \Phi_0 \rangle \, r_k^c \\ \nonumber 
&+ \sum_{klcd} \langle \Phi_i^a | \big[ \tilde{H}, \{ c^\dagger k d^\dagger l \} \big] | \Phi_0 \rangle \;
\frac{\sum_{me} \langle \Phi_{kl}^{cd} | \big[ \tilde{H}, \{ e^\dagger m \} \big] | \Phi_0 \rangle}
{\omega - \Delta_{kl}^{cd}} \, r_m^e \\ \nonumber 
&= \frac{1}{2} \sum_{kcd} r_{ik}^{cd} \langle \hat{a}k || cd \rangle - \frac{1}{2} \sum_{klc} r_{kl}^{ac} \langle kl || \hat{i}c \rangle 
+ \sum_{kc} r_{ik}^{ac} \tilde{F}_c^k + \sum_{kc} t_{ik}^{ac} \overline{F}_c^k \\
&- \frac{1}{2} \sum_{bckl} r_l^a t_{ik}^{cd} \langle lk || cd \rangle 
- \frac{1}{2} \sum_{bckl} r_i^b \, t_{kl}^{ac} \, \langle kl || bc \rangle 
+ \sum_b r_i^b \, \tilde{F}_b^a - \sum_j r_j^a \, \tilde{F}_i^j + \overline{F}_i^a
\label{eq:Aeff2} \end{align} 
Similar to Eq. \eqref{eq:cc2s2} for the reference state, Eq. \eqref{eq:Aeff1} can be solved without the need 
to store the double amplitudes $r_{ij}^{ab}$. In addition, the electron-repulsion integrals in Eqs. \eqref{eq:r2} 
and \eqref{eq:Aeff2} can be decomposed using the RI approximation according to Eq. \eqref{eq:ri}. The 
quantities $\overline{F}_a^i$ and $\overline{F}_i^a$ are defined as
\begin{equation} \label{eq:Fbar}
\overline{F}_i^a = \sum_{ck} r_k^c \, \langle ik || ac \rangle~, \qquad \overline{F}_a^i 
= \sum_{ck} r_k^c \, \langle ak || \hat{i}c \rangle~.
\end{equation}

\subsection{Computation of electron-attachment and ionization energies with the EOM-EA-CC2 and EOM-IP-CC2 
methods} \label{sec:eacc2}

An advantage of the EOM-CC approach is that it provides access not only to excited states, 
but also to ionized, electron-attached, and spin-flipped states. Ionized and electron-attached states 
can be generated using the operators
\begin{align} \label{eq:eomip}
R^\text{IP} &= R_1 + R_2 = \sum_i r_i  \, i + \frac{1}{2} \sum_{aij} r_{ij}^a \, a^\dagger i  j ~, \\
R^\text{EA} &= R_1 + R_2 = \sum_a r_a \, a^\dagger + \frac{1}{2} \sum_{abi} r_i^{ab} \, a^\dagger i b^\dagger 
\label{eq:eomea}
\end{align}
that have the same form in EOM-CC2 and EOM-CCSD theory. 

The EOM-EA-CC2 working equations can be derived from Eqs. \eqref{eq:r2} and \eqref{eq:Aeff2} by replacing 
$r_i^a$ by $r^a$ and $r_{ij}^{ab}$ by $r_j^{ab}$. This yields
\begin{align} \label{eq:eacc2d}
r_j^{ab} &= \frac{\sum_c r^c \langle \hat{a}\hat{b} || c\hat{j} \rangle}{\omega + \varepsilon_j - \varepsilon_a 
- \varepsilon_b }~, \\
\sigma^a &= (\mathbf{A}^\text{eff} \!\boldsymbol{\cdot} r)^a = \frac{1}{2} \sum_{kcd} r_{k}^{cd} 
\langle \hat{a}k || cd \rangle + \sum_{kc} r_{k}^{ac} \tilde{F}_c^k - \frac{1}{2} \sum_{bckl} r^b \, t_{kl}^{ac} \, 
\langle kl || bc \rangle + \sum_b r^b \, \tilde{F}_b^a~.  \label{eq:eacc2s}
\end{align}

Likewise, the EOM-IP-CC2 working equations can be derived from Eqs. \eqref{eq:r2} and \eqref{eq:Aeff2} by 
replacing $r_i^a$ by $r_i$ and $r_{ij}^{ab}$ by $r_{ij}^b$. This yields
\begin{align} \label{eq:ipcc2d}
r_{ij}^b &= - \frac{\sum_k r_k \langle k\hat{b} ||\hat{i}\hat{j} \rangle}{\omega + \varepsilon_i + 
\varepsilon_j - \varepsilon_b} ~, \\
\sigma_i &= (\mathbf{A}^\text{eff} \!\boldsymbol{\cdot} r)_i = - \frac{1}{2} \sum_{klc} r_{kl}^c \, 
\langle kl || \hat{i}c \rangle + \sum_{kc} r_{ik}^c \, \tilde{F}_c^k - \frac{1}{2} \sum_{bckl} r_l \, t_{ik}^{cd}\, 
\langle lk || cd \rangle - \sum_j r_j \, \tilde{F}_i^j ~.
\label{eq:ipcc2s}
\end{align}

\subsection{Spin scaling} \label{sec:spin}
Spin scaling was initially used to improve the accuracy of the MP2 energy\citep{grimme2003improved,
jung2004scaled} and later extended to CC2 theory.\citep{hellweg2008benchmarking} If 
$| \Phi_0 \rangle$ is a closed-shell wave function, the spin-scaled CC2 energy can be expressed as 
\begin{align} \label{eq:refsc}
E^\text{scaled}_\text{CC2} &= \sum_{iJaB} (ia|JB) \left[ t_i^a t_J^B + c_\text{os}\, t_{iJ}^{aB} \right] \\ \nonumber 
&+ \frac{1}{4} \sum_{ijab} \big[ (ia|jb) - (ib|ja) \big] \left[ t_i^a t_j^b - t_i^b t_j^a + c_\text{ss}\, t_{ij}^{ab} \right] \\
&+ \frac{1}{4} \sum_{IJAB} \big[ (IA|JB) - (IB|JA) \big] \left[ t_I^A t_J^B - t_I^B t_J^A + c_\text{ss}\, t_{IJ}^{AB} \right] 
\nonumber \end{align}
where uppercase and lowercase letters are used to distinguish $\alpha$ and $\beta$ spin orbitals. The 
constants $c_\text{os}$ and $c_\text{ss}$ are chosen as 6/5 and 1/3 in the SCS-CC2 method and as 1.3 
and 0 in the SOS-CC2 method. The same scaling factors $c_\text{os}$ and $c_\text{ss}$ are applied 
in a similar fashion to Eqs. \eqref{eq:cc2s2}, \eqref{eq:Aeff2}, \eqref{eq:eacc2s}, and \eqref{eq:ipcc2s} 
wherever the double amplitudes $t_{ij}^{ab}$ and $r_{ij}^{ab}$, $r_j^{ab}$, $r_{ij}^a$ appear. 

\subsection{Modifications for complex-energy CC2 methods} \label{sec:cxcc2}

All equations presented in Sections \ref{sec:ricc2} to \ref{sec:spin} remain valid if CC2 theory is built 
on a reference wave function $| \Phi_0 \rangle$ that includes a CAP or is represented in a basis set 
including complex-scaled functions. However, it is necessary to apply the c-product\citep{moiseyev2011non} to all 
expressions so that the bra state is not complex conjugated. With these modifications, the eigenvalues 
of the CC2 Jacobian (Eq. \eqref{eq:Aeff1}) become complex-valued and can be interpreted in terms 
of Eq. \eqref{eq:siegert}.

In the case of CAP calculations, all integrals over atomic orbitals (AOs) are real-valued and remain 
unchanged compared to the CAP-free case. The molecular orbital (MO) coefficients and all other 
wave function parameters, are, however complex-valued. In the case of CBF calculations, the AO 
integrals are complex-valued, but it is possible to use purely real-valued auxiliary basis sets without 
compromising accuracy so that $J_{PQ}$ from Eq. \eqref{eq:ri} is real-valued.\citep{hernandez2020resolution} 

\section{Implementation} \label{sec:imp}
\subsection{Code structure} \label{sec:code}

We added two independent implementations of real-energy RI-CC2 including the EOM-EE, EOM-EA, 
EOM-IP, and EOM-SF variants and complex-energy RI-CC2 including the EOM-EE and EOM-EA 
variants to the modules CCMAN2 and GMBPT of the Q-Chem software.\citep{Epifanovsky2021,
paran2022spin,paran2024performance,paran2024new}. These two implementations differ in 
their handling of the double excitation amplitudes and the electron-repulsion integrals, which has 
implications for the operation count and the memory requirements.

In the CCMAN2 module, which was originally developed for CCSD and EOM-CCSD calculations and 
uses the libtensor library for the handling of all tensors,\citep{epifanovsky2013new} we implemented the 
RI-CC2 method in the form of Eqs. \eqref{eq:cc2d} and \eqref{eq:cc2s1}, i.e., by removing terms from 
the RI-CCSD equations. Likewise, the working equations for the RI-EOM-CC2 methods are obtained 
from the corresponding RI-EOM-CCSD equations.\citep{epifanovsky2013general,camps2025complex} 
All equations are coded in spin-orbital form and the different spin cases are handled by libtensor. The 
EOM-CC2 eigenvalue equations are solved using the single-root or multiroot Davidson algorithms\citep{
davidson197514} that were implemented previously for EOM-CCSD calculations. 

The double amplitudes $t_{ij}^{ab}$ and $r_{ij}^{ab}$ are stored on disk in this CC2 implementation 
entailing $\mathcal{O}(N^4)$ memory requirements. Their contraction with the electron-repulsion 
integrals leads to an operation count of $\mathcal{O}(N^6)$, which means that the technical requirements 
are comparable to those of EOM-CCSD, limiting the range of application. However, this implementation 
supports point-group symmetry, resulting in computational savings for molecules that belong to higher 
point groups than $C_1$, and the EOM calculations can be directed towards the desired solution using 
all functionalities that were implemented previously for EOM-CCSD, such as user-defined guesses, 
root following, eigenvalue shifts, and pre-converging the single amplitude equations. 
 
Our second implementation uses the GMBPT module, which was originally developed for RI-MP2 
calculations.\citep{distasio2007improved} This implementation is more memory efficient and based 
on Eqs. \eqref{eq:cc2d2} and \eqref{eq:cc2s2}, partially transformed to the AO basis.\citep{
hattig2000cc2} Likewise, the implementation of the EOM methods in GMBPT is based on Eqs. 
\eqref{eq:Aeff2}, \eqref{eq:eacc2s}, and \eqref{eq:ipcc2s}, transformed to the AO basis. The 
$\hat{t}_{ij}^{ab}$ and $\hat{r}_{ij}^{ab}$ amplitudes are not stored; instead their contributions to 
the single amplitude equations are computed on the fly. 

This approach results in a scaling of the operation count as $\mathcal{O}(N^5)$ and memory 
requirements scaling as $\mathcal{O}(N^3)$ as no four-index quantity needs to be stored. 
Standard linear algebra operations are handled in the GMBPT module by the Armadillo library,\citep{
sanderson2018user} and the different spin cases are treated explicitly. This code structure makes 
it straightforward to apply scaling factors as discussed in Section \ref{sec:spin}. A difference 
between the two implementations lies in the treatment of complex-valued integrals. In CCMAN2, 
the real and imaginary parts are handled separately as real-valued objects, whereas in GMBPT, 
all integrals are treated directly as \textit{std::complex} objects. Moreover, in CCMAN2 the 
three-index electron-repulsion integrals arising from the RI approximation are precomputed and 
stored in memory, while they are recomputed and transformed by $t_i^a$ or $r_i^a$ in each CC 
or EOM-CC iteration in GMBPT. 

The pseudoeigenvalue equation from Eq. \eqref{eq:Aeff1} is solved with a modified Davidson 
algorithm, which is discussed further in Section \ref{sec:david}. For real-valued calculations in 
GMBPT, the convergence of the solution of Eq. \eqref{eq:Aeff1} is accelerated by extrapolation 
using direct inversion in the iterative subspace (DIIS)\citep{pulay1980convergence} after the residual error in the 
Davidson algorithm has dropped below some threshold, typically $10^{-5}$.\citep{hattig2000cc2,
paran2024new} In the complex-valued case, we observed faster convergence when this step is 
omitted and the solutions of Eq. \eqref{eq:Aeff1} are determined using solely using the algorithm 
from Section \ref{sec:david}. 

Also, we implemented user-defined guesses, root following, and eigenvalue shifts. However, the 
implementation in the GMBPT module currently does not exploit point-group symmetry, meaning 
all calculations are performed in the $C_1$ point group. In practice, we have found user-defined 
guesses to be more effective for calculations with GMBPT and eigenvalue shifts more useful for 
calculations with CCMAN2. If multiple roots are sought, they can be computed in GMBPT either 
sequentially or together using a multiroot algorithm similar to the implementation in CCMAN2.

Targeting a specific resonance state requires some prior knowledge of the resonance position or 
the character of the state. In this work, we found it useful to inspect the complex orbital energies 
from a CAP-HF or CBF-HF calculation to form suitable initial guesses for EA-CC2 calculations on 
temporary anions. Specifically, we usually selected orbitals with small imaginary energy for that 
purpose. If the choice of orbital was ambiguous, we resorted to the multiroot algorithm.

Further technical details concerning the implementation in the GMBPT module have been provided 
in Ref. \citep{paran2024new}. We finally note that our code was used to re-implement the CC2 
method with a stochastic RI approximation into the Q-Chem software.\citep{zhao2024stochasticproperties,zhao2024stochastic,zhao2025stochastic}

\subsection{Solution of the CC2 pseudoeigenvalue equation for excitation, attachment, or ionization energies}
\label{sec:david}

\begin{figure}[htbp] 
\includegraphics[width=1\textwidth]{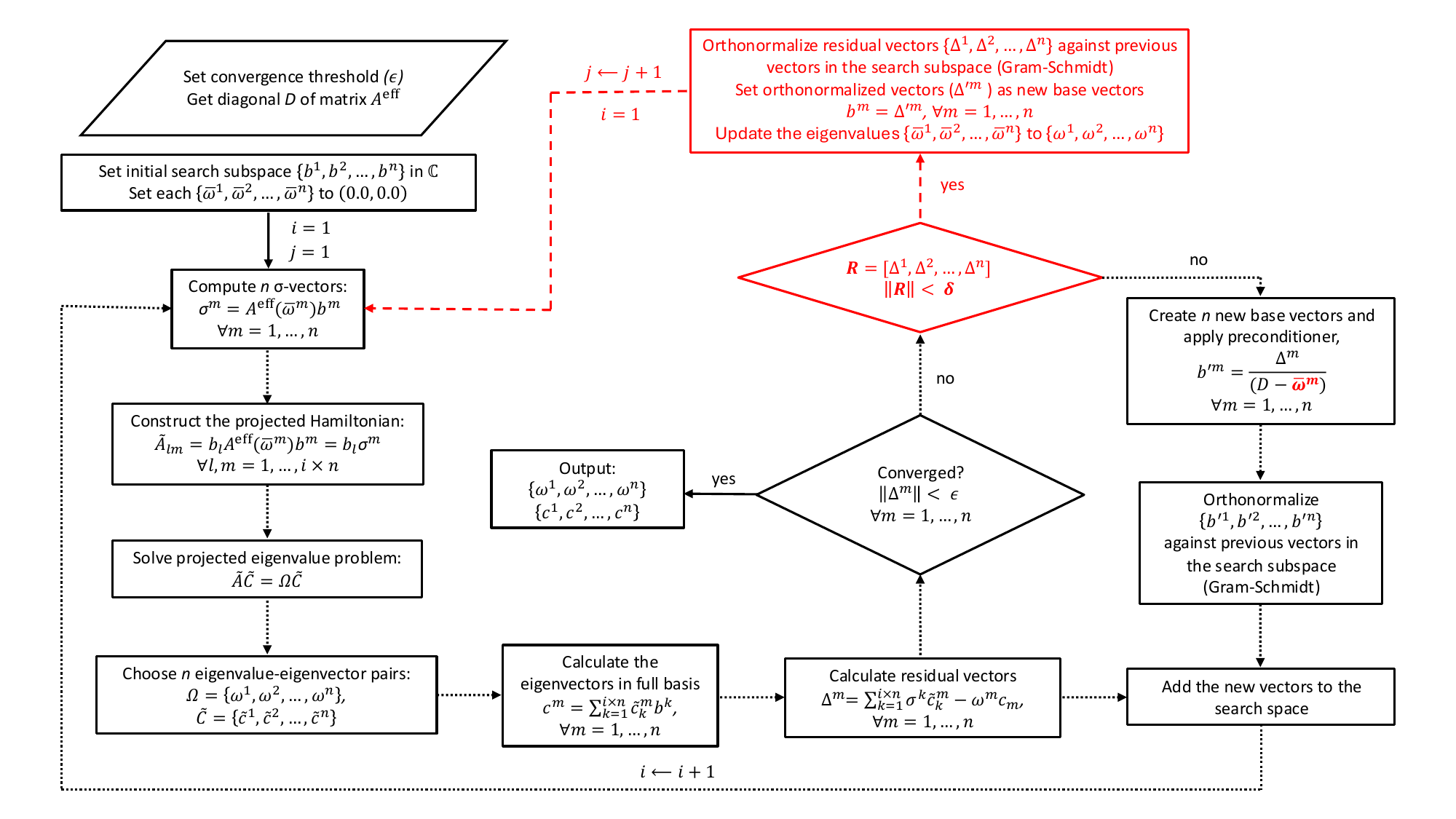}
\caption{Algorithm for solving the CC2 pseudoeigenvalue equation (Eq. \eqref{eq:Aeff1}). Macro iterations 
follow the dashed lines and micro iterations follow the dotted lines. Different from the standard Davidson 
algorithm are indicated in red.}
\label{fig:david}
\end{figure}

The algorithm by Davidson\citep{davidson197514} is commonly used to find a few solutions to the 
large-scale eigenvalue equations $\mathbf{A}\, C = \omega\, C$ occurring, for example, in EOM-CC, 
ADC, and CI theories, where a complete diagonalization is prohibitively expensive. A low-dimensional 
search subspace is constructed in an iterative manner and the original large matrix $\mathbf{A}$ is 
projected onto this subspace in each iteration. The projected eigenvalue equation $\tilde{\mathbf{A}}\, 
\tilde{C} = \tilde{\omega} \, \tilde{C}$ can be solved explicitly and an approximate eigenvalue 
$\tilde{\omega}$ and eigenvector $\tilde{C}$ are obtained. The quality of these approximate quantities 
is assessed by computing the residual vector $\Delta$ for the original large-scale eigenvalue equation. 
If the residual norm is sufficiently small, convergence is achieved. Otherwise, the subspace is expanded 
by an additional vector that is constructed from the residual using the zeroth-order Hamiltonian as 
preconditioner and orthogonalized against all current vectors.  

However, the CC2 pseudoeigenvalue equation (Eq. \eqref{eq:Aeff1}) is not a true eigenvalue equation 
because the Jacobian $\mathbf{A}^\text{eff}$ depends on the excitation energy $\omega$. To address 
this, we follow Ref. \citep{hattig2000cc2} and employ an algorithm that accounts for this energy 
dependence as shown in Fig. \ref{fig:david}. The key idea is to use two nested loops. In the inner loop, 
referred to as ``micro iterations'' in Ref. \citep{hattig2000cc2} and denoted by dotted lines in Fig. 
\ref{fig:david}, $\mathbf{A}^\text{eff}$ is held fixed, which renders Eq. \eqref{eq:Aeff1} a true eigenvalue 
equation. As soon as the residual has dropped below the threshold $\delta$, which is determined by 
comparing the current value for $\omega$ with that used to construct $\mathbf{A}^\text{eff}$, the inner 
loop is exited and $\mathbf{A}^\text{eff}$ is re-evaluated with the updated value for $\omega$. In the 
multiroot version of the algorithm, the elements of the projected Jacobian $\tilde{\mathbf{A}}$ are 
updated with the different approximate eigenvalues. At the start of each of these ``macro iterations'', 
denoted by dashed lines in Fig. \ref{fig:david}, the subspace is constructed anew, starting from the 
current best solution and discarding all previous subspace vectors. The macro iterations are continued 
until the residual has dropped below the convergence threshold $\epsilon$.

\section{Computational Details} \label{sec:cd}

The equilibrium geometries for the temporary anions C$_2$H$_4^-$, CH$_2$O$^-$ and HCOOH$^-$ 
were taken from Ref.\citep{benda2017communication} to enable a direct comparison with 
CAP-EOM-EA-CCSD results. For N$_2^-$, the bond length was set to 1.098\AA. In the case 
of uracil, naphthalene, and the two cyanonaphthalene isomers, we used the structures from 
Ref.~\citep{gayvert2022projected} to match the projected CAP-EOM-EA-CCSD reference data. 
The geometry of pyrene was optimized at the MP2/cc-pVTZ level. All structures are available 
in the supplementary material. We used the same equilibrium geometries for both CAP and 
CBF calculations to ensure consistency. The frozen core approximation was applied to uracil, 
naphthalene, the cyanonaphthalene isomers, and pyrene.

The CAP-RI-CC2 calculations were performed using basis sets consistent with those applied in 
Refs.~\citep{benda2017communication,gayvert2022projected}. For N$_2^-$, C$_2$H$_4^-$, 
CH$_2$O$^-$ and HCOOH$^-$, we used the aug-cc-pVDZ basis set augmented by three additional 
diffuse s- and p-shells on each non-hydrogen atom, denoted as aug-cc-pVDZ+3s3p. All calculations 
were also carried out with the aug-cc-pVTZ basis set, augmented with three additional s-, p-, and 
d-shells on each non-hydrogen atom. This basis is denoted as aug-cc-pVTZ+3s3p3d. The exponents 
of the additional shells are even tempered, wherein each exponent is obtained by dividing the preceding 
exponent by 2. For the larger molecules naphthalene, cyanonaphthalene, and pyrene, we only used 
the cc-pVDZ basis set, augmented by 2s, 5p, and 2d diffuse shells, resulting in cc-pVDZ+2s5p2d. 
This choice of basis was motivated by Ref.~\citep{ehara2012cap}, where the exponents for the 
s- and d-shells were obtained using a scaling factor of 2, and those for the p-shells were obtained 
using a scaling factor of 1.5. This same basis set was used for the CAP calculations on pyrene to 
ensure consistency.

\begin{figure}[htbp] \centering
\includegraphics[width=0.8\textwidth]{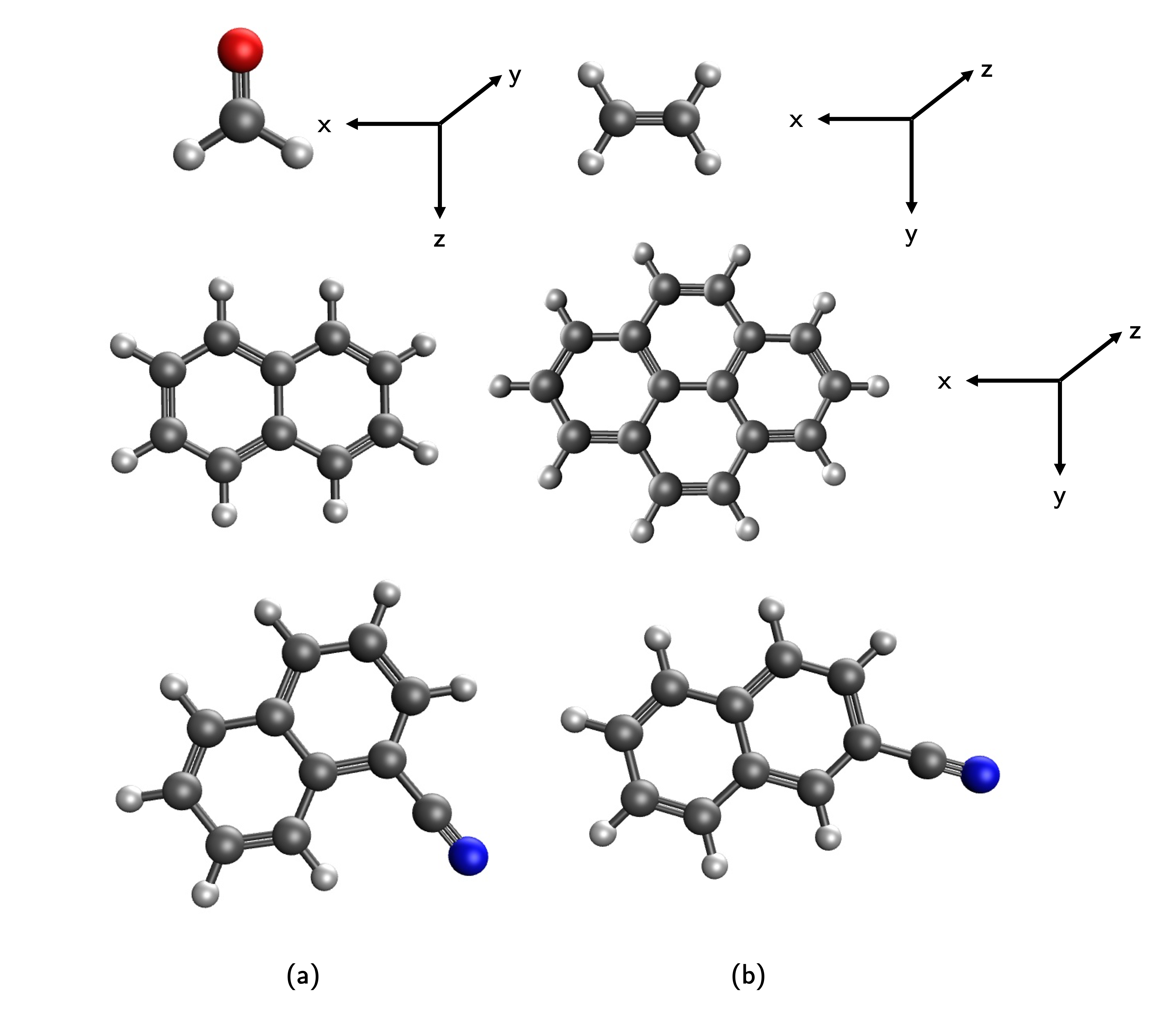}
\caption{Orientation of the molecules in our calculations. From top left to bottom right: formaldehyde, 
ethylene, naphthalene, pyrene, 1-cyanonaphthalene, and 2-cyanonaphthalene.}
\label{fig:orient}
\end{figure}

We used a quadratic box potential in all CAP-RI-CC2 calculations. CAP onsets were taken from 
Ref.~\citep{benda2017communication} for C$_2$H$_4^-$, CH$_2$O$^-$, and HCOOH$^-$ 
and from Ref.~\citep{gayvert2022projected} for N$_2^-$, naphthalene, and the cyanonaphthalene 
isomers. For pyrene, the CAP onset along each Cartesian coordinate was taken as the square root 
of the second moment of the electron density of the ground state, $\sqrt{\langle \alpha^2 \rangle}$ 
($\alpha = x, y, z$) computed at the CCSD/cc-pVDZ level. All CAP onset values are given in the 
supplementary material. The orientation of the molecules is displayed in Fig.~\ref{fig:orient}.

CBF-RI-CC2 calculations were carried out using the aug-cc-pVDZ+2s3p2d and aug-cc-pVTZ+3s3p3d 
basis sets. In these calculations, complex scaling was applied to the most diffuse Gaussian shells: 
one $s$- and $d$-, and two $p$-shells in the double-$\zeta$ basis; two shells of each angular momentum 
in the triple-$\zeta$ basis. We denote these basis sets as aug-cc-pVDZ+1s1p1d+\textit{1s2p1d} and 
aug-cc-pVTZ+1s1p1d+\textit{2s2p2d}, where the italicized part indicates the complex-scaled shells. 

The auxiliary basis was aug-cc-pVDZ-RIFIT or aug-cc-pVTZ-RIFIT, depending on the orbital basis 
set, for CAP and CBF calculations. No complex scaling was applied to the auxiliary basis functions, 
as previous RI-MP2 calculations have shown that this has a negligible effect on the results.\citep{
hernandez2020resolution}

An important computational advantage of the CBF approach is the similarity of the optimal complex 
scaling angles $\theta_\text{opt}$ at the EA-CC2 and EOM-EA-CCSD levels of theory. As illustrated 
in the supplementary material, the energy and width remain stable with deviations of the order of 
1 meV even if $\theta_\text{opt}$ can vary by 2--3$^\circ$. This suggests that the $\theta_\text{opt}$ 
value determined at the EA-CC2 level of theory can be used in EOM-EA-CCSD calculations, reducing 
the computational cost significantly. Notably, a similar trend does not apply to CAP calculations. Here, 
the optimal CAP strength $\eta_\text{opt}$ changes substantially when going from EA-CC2 to 
EOM-EA-CCSD, so that it is necessary to recalculate the full $\eta$-trajectory for each method. 

It is important to clarify how the irreducible representations were assigned to the anionic states of 
ethylene, formaldehyde, naphthalene, and pyrene. The CH$_2$O$^-$ molecule belongs to the 
$C_{2v}$ point group, whereas the other molecules belong to the $D_{2h}$ point group. For both 
of these point groups, the irreducible representations depend on the chosen molecular orientation. 
Throughout this work, all irreducible representations refer to the orientation shown in Fig.~\ref{fig:orient}, 
which was used in all calculations.

\section{Results and Discussion} \label{sec:res}
\subsection{Temporary anions of dinitrogen, ethylene, formaldehyde, and formic acid} \label{sec:res1}

To investigate the performance of complex-energy RI-EA-CC2, we first applied the new methods to a 
set of well-characterized temporary anions, N$_2^-$, C$_2$H$_4^-$, CH$_2$O$^{2-}$, HCOOH$^-$. 
These anions are small enough to enable a comparison to EOM-EA-CCSD. The vertical electron 
affinities and resonance widths computed with CAP and CBF versions of RI-EA-CC2 and EOM-EA-CCSD 
are shown in Table~\ref{tab:res1}. The corresponding optimal CAP strengths and complex-scaling 
angles are reported in the SI. 

\begin{table}[htbp] 
\setlength{\tabcolsep}{6pt} \renewcommand{\arraystretch}{0.80}
\caption{Vertical electron affinities of N$_2$, C$_2$H$_4$, CH$_2$O, and HCOOH and resonance 
widths of the corresponding temporary anions computed with RI-EA-CC2 and EOM-EA-CCSD 
combined with CAP or CBFs. All values in eV.} \vspace{0.2cm}
\begin{tabular}{rrrrrr} \hline\hline
Method & Basis set & N$_2^-$($^2\Pi_g$) & C$_2$H$_4^-$($^2$B$_{2g}$) & 
CH$_2$O$^-$($^2$B$_1$) & HCOOH$^-$($^2$A'') \\ \hline
\multicolumn{6}{c}{Vertical electron affinities computed with CAP} \\ \hline  
RI-EA-CC2 & aDZ+33 & --2.633 & --2.039 & --1.241 & --2.203 \\
EOM-EA-CCSD & aDZ+33 & --2.616 & --2.228\scite{benda2017communication} & 
--1.372\scite{benda2017communication} & --2.325\scite{benda2017communication}  \\
RI-EA-CC2 & aTZ+333 & --2.514 & --1.755 & --1.044 & --1.981 \\
SCS-RI-EA-CC2 & aTZ+333 & --2.803 & --2.042 & --1.425 & --2.319 \\
SOS-RI-EA-CC2 & aTZ+333 & --2.949 & --2.209 & --1.589 & --2.482 \\
EOM-EA-CCSD & aTZ+333 & --2.524\scite{gayvert2022projected} & --1.911 & --1.275 & --1.989 \\ \hline
\multicolumn{6}{c}{Vertical electron affinities computed with CBFs} \\ \hline
RI-EA-CC2 & aTZ+333* & --2.542 & --1.759 & --0.932 & --1.959 \\
SCS-RI-EA-CC2 & aTZ+333* & --2.849 & --2.091 & --1.281& --2.312 \\
SOS-RI-EA-CC2 & aTZ+333* & --3.006 & --2.262 & --1.461 & --2.491 \\
EOM-EA-CCSD & aTZ+333* & --2.552 & --1.993 & --1.106 & --2.151 \\ \hline 
Reference value &  & --2.32\scite{berman1983nuclear} & --1.76\scite{sanche1973electron} & 
--0.86\scite{burrow1976electron} & --1.73\scite{aflatooni2001temporary} \\ \hline\hline 
\multicolumn{6}{c}{Resonance widths computed with CAP} \\ \hline 
RI-EA-CC2 & aDZ+33 & 0.589 & 0.376 & 0.304 & 0.219 \\
EOM-EA-CCSD & aDZ+33 & 0.565 & 0.450\scite{benda2017communication} & 
0.353\scite{benda2017communication} & 0.252\scite{benda2017communication} \\
RI-EA-CC2 & aTZ+333 & 0.489 & 0.351 & 0.285 & 0.234 \\
SCS-RI-EA-CC2 & aTZ+333 & 0.589 & 0.491 & 0.376 & 0.293 \\
SOS-RI-EA-CC2 & aTZ+333 & 0.642 & 0.591 & 0.421 & 0.344 \\
EOM-EA-CCSD & aTZ+333 & 0.494\scite{gayvert2022projected} & 0.434 & 0.355 & 0.246 \\ \hline
\multicolumn{6}{c}{Resonance widths computed with CBFs} \\ \hline 
RI-EA-CC2 & aTZ+333* & 0.566& 0.536 & 0.465 & 0.372 \\
SCS-RI-EA-CC2 & aTZ+333* & 0.674& 0.718 & 0.658 & 0.470 \\
SOS-RI-EA-CC2 & aTZ+333* & 0.729& 0.812 & 0.771 & 0.527 \\
EOM-EA-CCSD & aTZ+333* & 0.559& 0.705 & 0.568 & 0.456 \\ \hline 
Reference value &  & 0.41\scite{berman1983nuclear} & 0.3--0.7\scite{sanche1973electron} & 
0.2--0.4\scite{burrow1976electron} & --- \\ \hline\hline 
\end{tabular} \vspace{0.3cm} \flushleft

aDZ+33 = aug-cc-pVDZ+3s3p; auxiliary basis for RI calculations is aug-cc-pVDZ-RIFIT. \\
aTZ+333 = aug-cc-pVTZ+3s3p3d; auxiliary basis for RI calculations is aug-cc-pVTZ-RIFIT. \\
aTZ+333* = aug-cc-pVTZ+1s1p1d+\textit{2s2p2d}; auxiliary basis for RI calculations is aug-cc-pVTZ-RIFIT.
\label{tab:res1} \end{table}

As shown in Table~\ref{tab:res1}, CAP-EA-CC2 and CBF-EA-CC2 both yield electron affinities that 
are about 0.1--0.2 eV smaller in magnitude than the corresponding EOM-EA-CCSD values, meaning 
that EA-CC2 places the temporary anions at absolute energies that are somewhat too low. This is 
consistent with the trend observed for bound anions of small molecules, which EA-CC2 also places 
at absolute energies that are too low by 0.1--0.2 eV, resulting in too big electron affinities.\citep{
ma2020approximate} Notably, the deviations between EA-CC2 and EOM-EA-CCSD observed for 
bound anions of larger molecules are substantially higher, typically amounting to 0.4--0.5 eV.\citep{
paran2024performance,shaalan2022accurate} We note that EA-CC2 agrees better with the 
experimental reference values for C$_2$H$_4^-$, CH$_2$O$^-$, and HCOOH$^-$ than 
EOM-EA-CCSD does, which we interpret as a fortuitous error cancellation.

A conspicuous exception to the general trend in Table \ref{tab:res1} is N$_2^-$; here the EA-CC2 
and EOM-EA-CCSD electron affinities deviate by no more than 0.02 eV. To check the validity of 
this result, we performed additional calculations for the $^2\Pi$ temporary anion state of CO$^-$, 
which is isoelectronic to N$_2^-$. CAP-EOM-EA-CCSD and CAP-EA-CC2 yield vertical electron 
affinities of --2.06 eV and --2.02 eV, respectively, in the aug-cc-pVTZ+3s3p3d basis set, which 
suggests that the different performance may be related to the electronic structure of these resonances.

The spin-scaled EA-CC2 methods both yield larger negative electron affinities than unmodified EA-CC2, 
the SCS variant leads to an increase of 0.3--0.4 eV and the SOS variant to an increase of about 0.5 eV. 
As a result, spin scaling does not improve the accuracy of EA-CC2 for the temporary anions in Table 
\ref{tab:res1}, but places the anions at too high absolute energies. This is different from bound valence 
anions of larger molecules for which electron affinities computed with SCS-EA-CC2 deviate by only 
0.07 eV from EOM-EA-CCSD.\citep{paran2024performance} We note that spin-scaling also does not 
improve the electron affinities of dipole-bound anions for which unmodified EA-CC2 yields very accurate 
results.\citep{paran2024performance}

The resonance widths computed with CAP-EA-CC2 and CBF-EA-CC2 exhibit smaller deviations of 
less than 0.1 eV from EOM-EA-CCSD. EA-CC2 consistently underestimates the resonance width, 
which we relate to a more severe underestimation at the HF level,\citep{white2015complex} where 
electron correlation is neglected completely. The spin-scaled EA-CC2 methods yield widths that are 
broader by 0.1--0.3 eV and again do not represent a clear improvement. N$_2^-$ is again a notable 
exception for which the widths computed with EA-CC2 and EOM-EA-CCSD differ by no more than 
0.025 eV. We also computed widths for CO$^-$, which yielded values of 0.87 eV and 0.88 eV with 
CAP-EOM-EA-CCSD and CAP-EA-CC2, respectively, confirming the trend observed for N$_2^-$. 

\subsection{Temporary anions of uracil} \label{sec:res2}

\begin{table}[htbp] 
\setlength{\tabcolsep}{9pt} \renewcommand{\arraystretch}{0.80}
\caption{Vertical electron affinities (EA) of uracil and resonance widths $\Gamma$ of the corresponding 
temporary anion states ($\pi_1^*$, $\pi_2^*$) computed with different methods. All values in eV.} \vspace{0.2cm}
\begin{tabular}{rrrrrr} \hline\hline
&  & \multicolumn{2}{c}{EA} & \multicolumn{2}{c}{$\Gamma$} \\
Method & Basis set & $\pi_1^*$ & $\pi_2^*$ & $\pi_1^*$ & $\pi_2^*$ \\ \hline
CBF-RI-EA-CC2 & aDZ+232* & --0.326 & --1.906 & 0.025 & 0.290 \\
CBF-SCS-RI-EA-CC2 & aDZ+232* & --0.758 & --2.353 & 0.045 & 0.307 \\
CBF-SOS-RI-EA-CC2 & aDZ+232* & --0.960 & --2.577 & 0.055 & 0.313 \\
CBF-RI-EOM-EA-CCSD & aDZ+232* & --0.745 & --2.329 & 0.053 & 0.332 \\ 
CBF-RI-EA-CC2 & aTZ+333* & --0.146 & --1.667 & 0.019 & 0.240 \\
CBF-SCS-RI-EA-CC2 & aTZ+333* & --0.558 & --2.131 & 0.042 & 0.270 \\
CBF-SOS-RI-EA-CC2 & aTZ+333* & --0.767 & 2.364 & 0.050 & 0.281 \\
CBF-RI-EOM-EA-CCSD$^a$ & aTZ+333* & --0.583 & --2.176 & 0.054 & 0.301 \\ \hline 
CAP-EOM-EA-CCSD (zero-order)$^b$ & DZ+252 & --0.731 & --2.284 & 0.05 & 0.232 \\
CAP-EOM-EA-CCSD (first-order)$^b$ & DZ+252 & --0.726 & --2.258 & 0.042 & 0.17 \\
CAP-SAC-CI (first-order)$^c$ & DZ+252 & --0.57 & --2.21 & 0.05 & 0.10 \\
R-Matrix$^d$ &  & --0.12  & --1.94 & 0.003 & 0.17 \\ \hline
Experiment$^e$ &  & --0.22 & --1.58 & -- & -- \\ \hline\hline
\end{tabular} \flushleft
DZ+252 = cc-pVDZ+2s5p2d. \\
aDZ+232* = aug-cc-pVDZ+1s1p1d+\textit{1s2p1d} and aug-cc-pVDZ-RIFIT as auxiliary basis. \\
aTZ+333* = aug-cc-pVTZ+1s1p1d+\textit{2s2p2d} and aug-cc-pVTZ-RIFIT as auxiliary basis. \\
$^a$ Complex scaling angle not optimized, instead taken from RI-EA-CC2 calculation.
$^b$ From Ref. \citep{gayvert2022projected}.
$^c$ From Ref. \citep{kanazawa2016low}.
$^d$ From Ref. \citep{dora2009r}.
$^e$ From Ref. \citep{aflatooni1998electron}
\label{tab:res2} \end{table}
As a first example of a somewhat larger molecule, we investigated the two lowest-lying $\pi^*$ temporary 
anion states of uracil (C$_4$H$_4$N$_2$O$_2$), both of which belong to the A'' representation of the 
C$_s$ point group. These results are shown in Table \ref{tab:res2}. Here, we used only the CBF technique 
combined with EA-CC2 and EOM-EA-CCSD; projected CAP-EOM-EA-CCSD and CAP-SAC-CI results 
have been taken from Refs. \citep{gayvert2022projected} and \citep{kanazawa2016low}, respectively. 

Table \ref{tab:res2} illustrates that the electron affinities computed with EA-CC2 are less negative by about 
0.4 eV than the EOM-EA-CCSD values, which is a bigger deviation than that observed for the smaller 
molecules in Table \ref{tab:res1}. Spin scaling improves the agreement of EA-CC2 with EOM-EA-CCSD 
significantly, with the SCS variant deviating by no more than 0.03 eV, which is in contrast with Table 
\ref{tab:res1}. The resonance widths computed with EA-CC2 are narrower than the EOM-EA-CCSD 
widths by about 0.05 eV, which is similar to the results from Table \ref{tab:res1}. Spin scaling improves 
the agreement, which is again in contrast to Table \ref{tab:res1}. 

Projected CAP-EOM-EA-CCSD is in very good agreement with CBF-EOM-EA-CCSD; the only 
significant deviation occurs for the resonance width of the $\pi_2^*$ state, which is about 0.1~eV 
narrower when computed with CAP as compared to the CBF result. Notably, the electron affinity 
computed with CAP-SAC-CI deviates more substantially from this value, which must be caused 
by different CAP parameters or a different basis set used for the calculation, as the EOM-CCSD 
and SAC-CI Jacobian matrices have identical eigenvalues. 

Similar to the results in Table \ref{tab:res1}, the EA-CC2 results for the electron affinities of both 
anion states of uracil are in better agreement with the experimental reference values than the 
EOM-EA-CCSD results. In the triple-$\zeta$ basis, the deviation of EA-CC2 from the experimental 
values amounts to less than 0.1 eV. 

\subsection{Anions of naphthalene and cyanonaphthalene} \label{sec:res3}

Next, we investigated the lowest three $\pi^*$ anion states of naphthalene (C$_{10}$H$_8$) as well as 
1- and 2-cyanonaphthalene (C$_{11}$NH$_7$). These states belong to the B$_{2g}$, B$_{3g}$, and 
B$_{1u}$ representations, respectively, of the D$_{2h}$ point group for naphthalene and to the A'' 
representation of the C$_s$ point group for the cyanonaphthalenes. Notably, the lowest-lying state is 
bound for the cyanonaphthalenes, while unsubstituted naphthalene does not support a bound anion. 
This can be related to the electron-withdrawing character of the cyano group. The cyanonaphthalene 
molecules were recently detected in the Taurus Molecular Cloud-1,\citep{mcguire2021detection} which 
motivates the renewed interest in the temporary anion states of these molecules. 

The anions of cyanonaphthalene are at the current practical limit for a computational characterization 
using complex-energy EOM-EA-CCSD in appropriately large basis sets. The projected CAP approach 
offers a possible solution in this regard, but the application of CBFs is challenging. This motivates using 
EA-CC2 for the present study.

Our results for naphthalene and the cyanonaphthalenes are shown in Tables \ref{tab:res3} and \ref{tab:res4}, 
respectively. We first computed the vertical electron affinities and resonance widths of naphthalene 
and its cyano derivatives using CAP-EA-CC2 and compared them to the CAP-EOM-EA-CCSD results 
from Ref.~\citep{gayvert2022projected}, employing the same double-$\zeta$ basis set. Since the 
results from Ref. \citep{gayvert2022projected} were obtained with a projected smoooth Voronoi CAP, 
we recomputed the CAP-EOM-EA-CCSD energy of unsubstituted naphthalene using a box CAP. As 
Table \ref{tab:res3} illustrates, the form of the CAP does not affect the results significantly, which is 
consistent with previous findings.\citep{SommerfeldEhara2015_VoronoiCAP,GyamfiJagau2024_CAPstrategy} 

We then computed the resonances energies and widths with the CBF approach, employing 
double-$\zeta$ and triple-$\zeta$ basis sets. While we used CBF-EA-CC2 in both basis sets, we 
used CBF-EOM-EA-CCSD only in the double-$\zeta$ basis set. In addition, a CBF-EOM-EA-CCSD 
calculation was carried out for naphthalene in the triple-$\zeta$ basis using the optimal complex 
scaling angle $\theta_\text{opt}$ from the corresponding CBF-EA-CC2 calculation. This is possible 
because $\theta_\text{opt}$ values from EA-CC2 and EOM-EA-CCSD calculations are very similar 
as discussed in Section \ref{sec:cd}. For example, for the $\pi_2^*$ state of naphthalene, 
$\theta_\text{opt}$ values from CBF-EA-CC2 and CBF-EOM-EA-CCSD calculations differ by 4$^\circ$ 
in the double-$\zeta$ basis. If the CBF-EOM-EA-CCSD energy is evaluated at the $\theta_\text{opt}$ 
from CBF-EA-CC2, the result changes by less than 1 meV for both the real and the imaginary part. 
This approach saves a lot of compute time: One CBF-RI-EOM-EA-CCSD calculation in the 
aug-cc-pVTZ+1s1p1d+\textit{2s2p2d} basis set (734 functions) took approximately 120 hours 
using 16 threads of an Intel(R) Xeon(R) Gold 6334 processor (3.6 GHz), while the corresponding 
CBF-RI-EA-CC2 calculation only took 4 hours on the same machine.

\begin{table}[htb] \centering
\setlength{\tabcolsep}{9pt} \renewcommand{\arraystretch}{0.80}
\caption{Vertical electron affinities of naphthalene and resonance widths of the corresponding temporary 
anion states computed with RI-EA-CC2 and EOM-EA-CCSD combined with CAP or CBFs. 
All values in eV.} \vspace{0.2cm}
\begin{tabular}{rrrrrrrr} \hline\hline
 &  & \multicolumn{3}{c}{Vertical electron affinity} &\multicolumn{3}{c}{Resonance width} \\
Method & Basis set & $\pi_1^*$  & $\pi_2^*$ & $\pi_3^*$ & $\pi_1^*$  & $\pi_2^*$ & $\pi_3^*$ \\ \hline
CAP-RI-EA-CC2$^a$ & DZ+252 & --0.252 & --0.934 & --1.687 & 0.004 & 0.032 & 0.443 \\
CAP-EOM-EA-CCSD$^a$ & DZ+252 & --0.673 & --1.327 & --2.120 & 0.014 & 0.067 & 0.442 \\
CAP-EOM-EA-CCSD$^b$ & DZ+252 & --0.680 & --1.343 & --2.198 & 0.029 & 0.059 & 0.499 \\ \hline
CBF-RI-EA-CC2 & aDZ+232* & --0.265 & --0.962 & --1.669 & 0.003 & 0.038 & 0.449 \\
CBF-SCS-RI-EA-CC2 & aDZ+232* & --0.594 & --1.229 & --2.021 & 0.013 & 0.049 & 0.522 \\
CBF-SOS-RI-EA-CC2 & aDZ+232* & --0.761 & --1.356 & --2.201 & 0.021 & 0.053 & 0.569 \\
CBF-RI-EOM-EA-CCSD & aDZ+232* & --0.683 & --1.353 & --2.187 & 0.022 & 0.062 & 0.662 \\
CBF-RI-EA-CC2 & aTZ+333* & --0.074 & --0.756 & --1.487 & --0.001 & 0.028 & 0.379 \\
CBF-SCS-RI-EA-CC2 & aTZ+333* & --0.414 & --1.040 & --1.847 & 0.007 & 0.041 & 0.468 \\
CBF-SOS-RI-EA-CC2$^c$ & aTZ+333* & --0.583 & --1.181 & --2.032 & 0.013 & 0.042 & 0.515 \\
CBF-RI-EOM-EA-CCSD$^c$ & aTZ+333* & --0.536 & --1.203 & --2.051 & 0.013 & 0.053 & 0.622 \\ \hline
Experiment$^d$ &  & --0.19 & --0.90  & --1.67 & --- & --- & --- \\ \hline\hline
\end{tabular} \flushleft
DZ+252 = cc-pVDZ+2s5p2d and cc-pVDZ-RIFIT as auxiliary basis set. \\
aDZ+232* = aug-cc-pVDZ+1s1p1d+\textit{1s2p1d} and aug-cc-pVDZ-RIFIT as auxiliary basis set. \\
aTZ+333* = aug-cc-pVTZ+1s1p1d+\textit{2s2p2d} and aug-cc-pVTZ-RIFIT as auxiliary basis set. \\
$^a$ Computed using box CAP.
$^b$ From Ref. \citep{gayvert2022projected}, computed using projected smooth Voronoi CAP.
$^c$ Complex scaling angle not optimized, instead taken from RI-EA-CC2 calculation. 
$^d$ From Ref. \citep{burrow1987electron}.
\label{tab:res3}
\end{table}

\begin{table} \centering
\setlength{\tabcolsep}{9pt} \renewcommand{\arraystretch}{0.80}
\caption{Vertical electron affinities of cyanonaphthalene and resonance widths of the corresponding 
temporary anion states computed with RI-EA-CC2 and EOM-EA-CCSD combined with CAP or CBFs. 
All values in eV.} \vspace{0.2cm}
\begin{tabular}{rrrrrrr} \hline\hline
 &  & \multicolumn{3}{c}{Vertical electron affinity} &\multicolumn{2}{c}{Resonance width} \\
Method & Basis set & $\pi_1^*$  & $\pi_2^*$ & $\pi_3^*$ & $\pi_2^*$ & $\pi_3^*$ \\ \hline
 &  & \multicolumn{5}{c}{1-Cyanonaphthalene} \\ \hline
CAP-RI-EA-CC2$^a$ & DZ+252 & 0.678 & --0.315 & --0.835 & 0.005 & 0.137 \\
CAP-EOM-EA-CCSD$^b$ & DZ+252 & 0.226 & --0.760 & --1.398 & 0.033 & 0.226 \\ \hline
CBF-RI-EA-CC2 & aDZ+232* & 0.654& --0.337 & --0.895 & 0.005 & 0.137 \\
CBF-SCS-RI-EA-CC2 & aDZ+232* & 0.305 & --0.620 & --1.254 & 0.011 & 0.177 \\
CBF-SOS-RI-EA-CC2 & aDZ+232* & 0.128 & --0.762 & --1.436 & 0.022 & 0.203 \\
CBF-RI-EOM-EA-CCSD & aDZ+232* & 0.207 & --0.962 & --1.669 & 0.038 & 0.449 \\
CBF-RI-EA-CC2 & aTZ+333* & 0.859 & --0.129 & --0.701 & 0.002 & 0.106 \\
CBF-SCS-RI-EA-CC2$^c$ & aTZ+333* & 0.497 & --0.425 & --1.074 & 0.007 & 0.146 \\
CBF-SOS-RI-EA-CC2$^c$ & aTZ+333* & 0.313 & --0.574 & --1.263 & 0.015 & 0.169 \\ \hline 
Experiment$^d$ &  \multicolumn{2}{r}{0.68 $\pm$ 0.10} \\ \hline 
 &  & \multicolumn{5}{c}{2-Cyanonaphthalene} \\ \hline
CAP-RI-EA-CC2$^a$ & DZ+252 & 0.602 & --0.068 & --0.788 & 0.003 & 0.379 \\
CAP-EOM-EA-CCSD$^b$ & DZ+252 & 0.146 & --0.502 & --1.582 & 0.028 & 0.376 \\ \hline 
CBF-RI-EA-CC2 & aDZ+232* & 0.577 & --0.089 & --1.030 & --0.004 & 0.316 \\
CBF-SCS-RI-EA-CC2 & aDZ+232* & 0.233 & --0.384 & --1.385 & 0.013 & 0.385 \\
CBF-SOS-RI-EA-CC2 & aDZ+232* & 0.058 & --0.535 & --1.568 & 0.019 & 0.425 \\
CBF-RI-EOM-EA-CCSD & aDZ+232* & 0.127 & --0.512 & --1.569 & 0.024 & 0.499 \\
CBF-RI-EA-CC2 & aTZ+333* & 0.785 & --0.132 & --0.840  & 0.079 & 0.280 \\
CBF-SCS-RI-EA-CC2$^c$ & aTZ+333* & 0.427 & --0.138 & --1.213  & 0.084 & 0.332 \\
CBF-SOS-RI-EA-CC2$^c$ & aTZ+333* & 0.246 & --0.141 & --1.400  & 0.088 & 0.368 \\ \hline
Experiment$^d$ & \multicolumn{2}{r}{0.65 $\pm$ 0.10} \\ \hline\hline 
\end{tabular} \flushleft

DZ+252 = cc-pVDZ+2s5p2d and cc-pVDZ-RIFIT as auxiliary basis set. \\
aDZ+232* = aug-cc-pVDZ+1s1p1d+\textit{1s2p1d} and aug-cc-pVDZ-RIFIT as auxiliary basis set. \\
aTZ+333* = aug-cc-pVTZ+1s1p1d+\textit{2s2p2d} and aug-cc-pVTZ-RIFIT as auxiliary basis set. \\
$^a$ Computed using box CAP.
$^b$ From Ref. \citep{gayvert2022projected}, computed using projected smooth Voronoi CAP.
$^c$ Complex scaling angle not optimized, instead taken from RI-EA-CC2 calculation. 
$^d$ From Ref. \citep{heinis1993electron}, adiabatic value. 
\label{tab:res4}
\end{table}

At the EA-CC2 level of theory, the anions are on average 0.4--0.6 eV lower in energy than at the 
EOM-EA-CCSD level, with the $\pi_3^*$ state exhibiting the largest deviations. The CAP calculations 
in the double-$\zeta$ basis for the $\pi^*_3$ state of 2-cyanonaphthalene stand out with the largest 
deviation of 0.8 eV between EA-CC2 and EOM-EA-CCSD. Spin scaling raises the EA-CC2 energies 
of the anion states by about 0.3--0.6 eV and brings them closer to the EOM-EA-CCSD values. In 
general, the SOS approach leads to bigger shifts and better agreement with EOM-EA-CCSD, deviating 
by more than 0.1 eV only in two cases, the $\pi_2^*$ and $\pi_3^*$ states of 1-cyanonaphthalene. 
Notably, we observed a somewhat better performance of the SCS approach instead in our earlier 
study on bound valence anions.\citep{paran2024performance}

Table \ref{tab:res4} shows that all computational methods agree with the experimental result that 
1-cyanonaphthalene has a somewhat larger electron affinity than 2-cyanonaphthalene. However, 
Tables \ref{tab:res3} and \ref{tab:res4} also demonstrate that the EA-CC2 energies for the $\pi^*$ 
resonances of naphthalene and for the bound anions of the cyanonaphthalenes are in considerably 
better agreement with experiment than the EOM-EA-CCSD energies. We emphasize that the electron 
affinities of the cyanonaphthalenes obtained from gas-phase ion-molecule equilibrium constants should 
be interpreted as adiabatic values.\citep{heinis1993electron} Likewise, the data from electron transmission 
spectroscopy for the naphthalene temporary anions have been interpreted as adiabatic electron 
affinities as well,\citep{lyapustina2000solvent} even though they were originally presumed to be vertical values.\citep{
burrow1987electron} This ambiguity complicates the comparison with our computed vertical electron 
affinities. 

As concerns the resonance width, we find that EA-CC2 yields lower values than EOM-EA-CCSD, 
which is partially corrected by spin scaling, similar to what we discussed in Sections \ref{sec:res1} 
and \ref{sec:res2}. The deviations between the methods are small for the narrow $\pi_1^*$ and 
$\pi_2^*$ resonances but can amount to more than 0.2 eV for the broader $\pi_3^*$ resonances. 
It is also worth noting that the CBF-EA-CC2 method struggles with the very narrow resonance widths 
of the $\pi_1^*$ state of naphthalene and the $\pi_2^*$ state of 2-cyanonaphthalene, where it yields 
unphysical positive imaginary energies, i.e., negative decay widths in some basis sets. This may be 
caused by the RI approximation similar to what we found in previous investigations of autoionizing 
Rydberg states,\citep{camps2025complex} but it may also indicate that a higher-level treatment of 
correlation is needed to describe the decaying character of these states. Notably, the spin-scaled 
EA-CC2 methods do not exhibit this problem. 

The comparison between CAP and CBF methods illustrates very good agreement with deviations 
below 0.1 eV for the electron affinities and even lower deviation for the resonance widths in most 
cases. We note that, with both techniques, the resonance states can in some cases not be identified 
from their energy alone. Rather, it is necessary to analyze the EOM-CC eigenvectors: While resonance 
states show significant contributions from attachment to different orbitals, this is not the case for 
pseudocontinuum states, which are usually dominated to $>$99\% by attachment to one single orbital. 
Also, the determination of $\theta_\text{opt}$ was ambiguous in some CBF calculations as there were 
two minima in $|dE/d\theta|$. However, the corresponding energies and widths in Tables \ref{tab:res3} 
and \ref{tab:res4} would change by less than 0.002 eV if they were evaluated at the respective other 
$\theta$-values.

\subsection{Anions of pyrene} \label{sec:res4}

\begin{figure}[htbp] \centering
\includegraphics[width=0.65\textwidth]{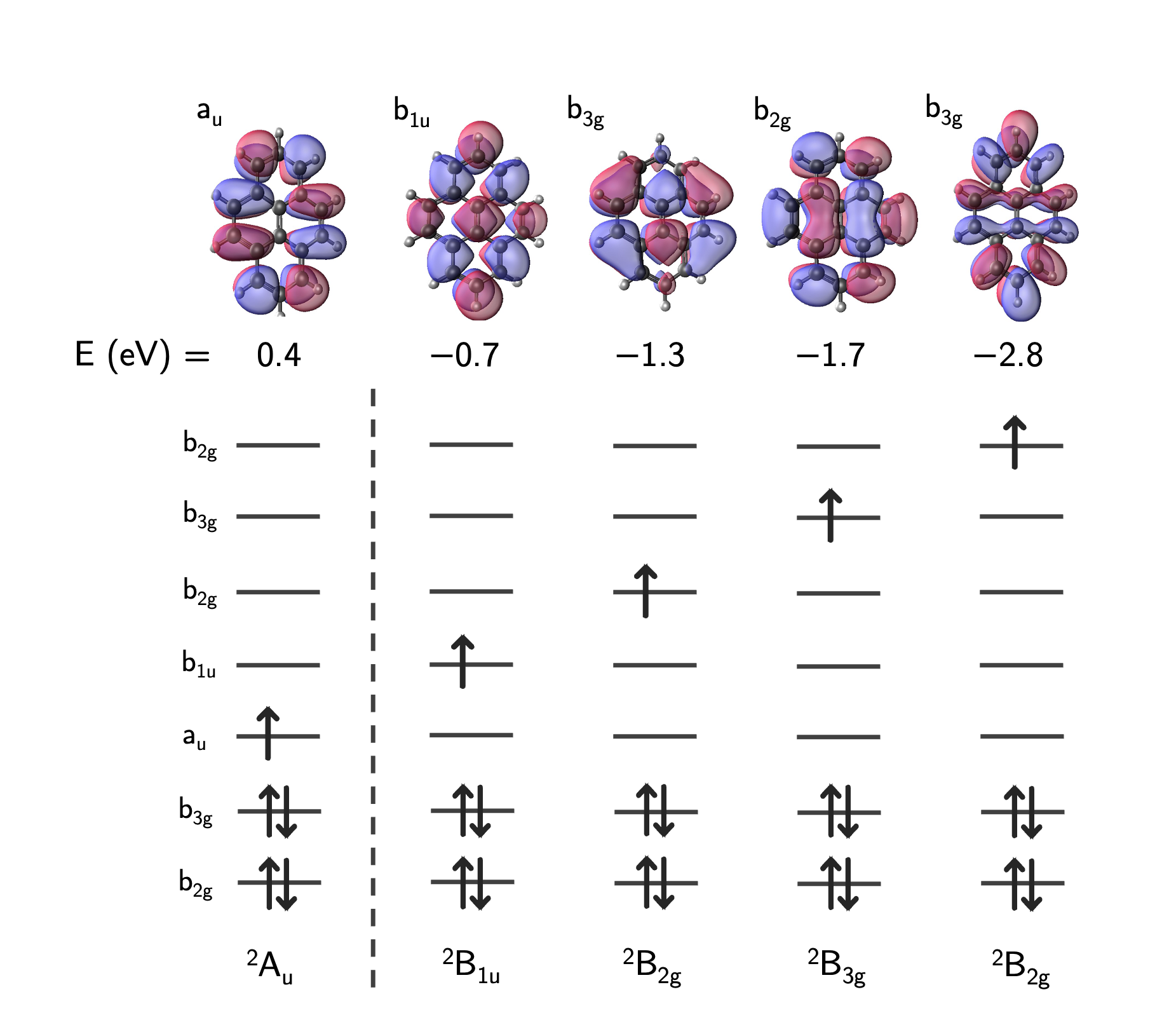}
\caption{Electronic configurations of the low-lying anion states of pyrene and corresponding Hartree-Fock 
orbitals computed in the cc-pVTZ basis without f-shells on carbon atoms and d-shells on hydrogen atoms. 
Electron affinities were extracted from 2D photoelectron spectroscopy. Figure adapted from 
Ref.~\citep{lietard2021temporary} with permission.}
\label{fig:pyrene} \end{figure}

\begin{table}[htbp] \centering
\renewcommand{\arraystretch}{0.80} \setlength{\tabcolsep}{6pt}
\caption{Vertical electron affinities of pyrene and resonance widths of the corresponding temporary 
anion states computed with RI-EA-CC2 and EOM-EA-CCSD combined with CAP or CBFs. Results 
from R matrix calculations and experimental values from photodetachment spectroscopy are also shown. 
All values in eV.} \vspace{0.2cm}
\begin{tabular}{rrrrrrrrrr} \hline\hline 
 & \multicolumn{5}{c}{Vertical electron affinity} & \multicolumn{4}{c}{Resonance width} \\
Method & $^2$A$_u$ & $^2$B$_{1u}$ & $^2$B$_{2g}$ & $^2$B$_{3g}$ & $^2$B$_{2g}$ & 
$^2$B$_{1u}$ & $^2$B$_{2g}$ & $^2$B$_{3g}$ & $^2$B$_{2g}$ \\ \hline  
CAP-RI-EA-CC2$^a$ & 0.501 & --0.239 & --0.592 & --0.878 & --2.583 & 0.003 & 0.016 & 0.203 & 0.052 \\
CBF-RI-EA-CC2$^b$ & 0.678 & --0.057 & --0.427 & --0.870 & --2.389 & 0.003 & 0.019 & 0.289 & 0.095 \\
CBF-SCS-RI-EA-CC2$^b$ & 0.333 & --0.335 & --0.840 & --1.276 & --2.728 & 0.011 & 0.056 & 0.388 & 0.113 \\
CBF-SOS-RI-EA-CC2$^b$ & 0.156 & --0.507 & --1.052 & --1.481 & --2.898 & 0.014 & 0.072 & 0.414 & 0.121 \\ \hline 
EOM-EA-CCSD$^c$ & 0.17 & --0.83 & --1.43 & --1.93 & --3.23 & --- & --- & --- & --- \\
R-Matrix$^d$ & --- & --0.5 & --1.1 & --1.6 & --3.1 & 0.06 & 0.28 & 0.63 & 0.14 \\ \hline 
Experiment$^e$ & 0.4 & --0.7 & --1.3 & --1.7 & --2.8 & --- & --- & --- & --- \\ \hline\hline
\end{tabular} \flushleft
$^a$ Basis set: cc-pVDZ+2s5p2d, auxiliary basis set: cc-pVDZ-RIFIT. \\
$^b$ Basis set: aug-cc-pVTZ+1s1p1d+\textit{2s2p2d}, auxiliary basis set: aug-cc-pVTZ-RIFIT. \\
$^c$ From Ref. \citep{lietard2021temporary}. Computed in the cc-pVTZ-f/d basis set. \\
$^d$ From Ref. \citep{singh2020electron}. Results shifted by 1.0 eV as motivated in 
Ref. \citep{lietard2021temporary}. \\
$^e$ From Ref. \citep{lietard2021temporary}. Adiabatic value obtained from two-dimensional 
photoelectron spectroscopy. 
\label{tab:pyr} \end{table}

As a final example, we investigated anion states of pyrene (C$_{16}$H$_{10}$), which can be viewed 
as a larger homologue of naphthalene. This planar, fully $\pi$-conjugated molecule supports one bound 
anionic state ($^2$A$_u$). Pyrene offers an ideal test case due to the availability of recent experimental 
data from photodetachment spectroscopy.\citep{lietard2021temporary} To elucidate the spectrum, 
EOM-EA-CCSD calculations in the cc-pVTZ basis set were carried out in Ref. \citep{lietard2021temporary}. 
These calculations neglect the continuum and give no information on the decay width, but they provide 
a qualitative picture of the electronic structure of the four low-lying $\pi^*$ shape resonances of pyrene 
as shown in Fig. \ref{fig:pyrene}. The energies computed with EOM-EA-CCSD/cc-pVTZ agree within 
0.1--0.4 eV with the photodetachment spectrum. Part of this disagreement can be ascribed to structural 
relaxation, which in Ref. \citep{lietard2021temporary} was estimated to account for less than 0.2 eV 
for the bound anion state. 

Our complex-energy EA-CC2 results are reported in Table~\ref{tab:pyr} and show a consistent pattern 
for all electronic states and methods. All EA-CC2 variants qualitatively agree on the order and the energetic 
spacing of the states. R-matrix results agree on the order and spacing as well,\citep{singh2020electron} 
but these calculations fail to identify a bound anion state and the energies of the unbound states need 
to be lowered by 1.0 eV to align with the experimental data.\citep{lietard2021temporary} 

Spin-scaling lowers the EA-CC2 energies of the bound and temporary anion states alike and the 
SOS-EA-CC2 result for the bound $^2$A$_u$ state is in very good agreement with the EOM-EA-CCSD 
energy from Ref. \citep{lietard2021temporary}. Based on the results from Sections \ref{sec:res1} to 
\ref{sec:res3}, we consider the SOS-EA-CC2 results to be our best estimates of the vertical electron 
affinities of the anion states. Indeed, SOS-EA-CC2 provides the closest estimates of the experimental 
values for the energies of the unbound states. For the bound anion state, this is not the case, but we 
note that the experimental value of 0.4 eV represents an adiabatic electron affinity, while our results 
refer to vertical energy differences. 

As concerns the resonance widths, all EA-CC2 variants yield a narrow width of less than 15 meV 
for the lowest resonance state ($^2$B$_{1u}$), corresponding to a lifetime of more than 40 fs. For 
the two next higher resonances ($^2$B$_{2g}$ and $^2$B$_{3g}$), the width grows progressively, 
but the fourth resonance ($^2$B$_{2g}$) is considerably narrower again. This somewhat unusual 
pattern is consistent with results from R-matrix calculations.\citep{singh2020electron} Also, we note 
that CBF-EA-CC2 consistently yields larger widths than CAP-EA-CC2, while the R-matrix method 
delivers even larger widths.

\section{Summary and Conclusions} \label{sec:con}

We have presented two independent implementations of the RI-EA-CC2 and RI-EE-CC2 methods 
for complex-valued Siegert energies. Our implementations can use complex absorbing potentials 
or, alternatively, complex basis functions and include spin-scaled CC2 variants as well. In addition, 
the EE, EA, IP, and SF variants of RI-CC2 have been implemented for real-valued energies. Our 
first implementation is based on the corresponding EOM-CCSD codes,\citep{zuev2014complex,
camps2025complex} whereas our second implementation avoids the storage of all double excitation 
amplitudes,\citep{hattig2000cc2} making it considerably more efficient in terms of memory and 
operation count. 

We benchmarked the performance of the new complex-energy EA-CC2 methods in appropriately 
augmented double-$\zeta$ and triple-$\zeta$ basis sets for a test set of 18 temporary anion states 
in molecules of varying size, including 4 states of pyrene (C$_{16}$H$_{10}$) as largest example. 
We find that EA-CC2 in general places the anion states at lower energies than EOM-EA-CCSD, 
i.e., EA-CC2 yields electron affinities that are less negative. The deviation between the methods 
does not exceed 0.04 eV for the smallest examples in our test set (N$_2^-$ and CO$^-$), but 
amounts to about 0.5 eV for larger molecules. Spin scaling raises the anion energies so that the 
agreement with EOM-EA-CCSD becomes significantly better for the anions of the larger molecules. 
Notably, all these trends are in agreement with previous findings for bound anions.\citep{
ma2020approximate,shaalan2022accurate,paran2024performance}

We observe a seemingly better agreement of experimental resonance energies with EA-CC2 
results than with EOM-EA-CCSD results. However, this comparison with experiment should be 
treated with caution as it is not always clear whether the available data refer to vertical or adiabatic 
electron affinities. For the bound anions of pyrene and cyanonaphthalene, for which it is clear that 
the experimental values are adiabatic, it is obvious that the good performance of EA-CC2 is at least 
partially due to a cancelation with the reorganization energy caused by structural relaxation. For the 
resonance width, EA-CC2 consistently yields lower values than EOM-EA-CCSD, which agrees with 
previous findings that HF theory yields even narrower widths. Also for the width, spin scaling improves 
the agreement with EOM-EA-CCSD. 

In general, CAP and CBF calculations are in very good agreement for resonance energies and 
widths. However, an advantage of the CBF approach is that the optimal value for the complex scaling 
angle is very similar at the EA-CC2 and EOM-EA-CCSD levels of theory, which makes it possible to 
determine it with EA-CC2 and then run a single EOM-EA-CCSD calculation at this value.

Our work illustrates that EA-CC2 can be readily applied to temporary anion states that are too big 
for a complex-energy EOM-EA-CCSD treatment such as those of pyrene. Also, for cases for which 
EOM-EA-CCSD calculations are still feasible, EA-CC2 offers a drastic speedup. For example, 
an EOM-EA-CCSD calculation on naphthalene in a triple-$\zeta$ basis took 120 hours, but the 
corresponding EA-CC2 calculation could be completed within 4 hours and delivered, when spin 
scaling was applied, the same energy and width within less than 0.1 eV. Therefore, we expect 
that our implementation will be useful for the investigation of temporary anions in larger molecules. 

To enable a more comprehensive characterization of temporary anions beyond energies and widths, 
we consider it worthwhile to extend the complex-energy EA-CC2 methods to molecular properties 
and analytic gradients similar to what was done for EOM-EA-CCSD.\citep{jagau2016characterizing,
benda2017communication,benda2018understanding,mondal2025analytic} Also, we consider it promising to apply complex-valued 
CC2 methods to other resonances besides temporary anions. Specifically, this includes IP-CC2 
calculations for Auger decay of core-ionized states or intermolecular Coulombic decay of ionized 
clusters, and EE-CC2 calculations for autoionization of superexcited Rydberg states. 

\section*{Acknowledgments}

The authors thank Dr. Valentina Parravicini, Dr. Anthuan Ferino P\'erez, and Simen Camps for useful 
discussions on the implementation. T.-C.J. gratefully acknowledges funding from the European Research 
Council (ERC) under the European Union's Horizon 2020 research and innovation program (grant 
agreement no. 851766) and from the KU Leuven internal funds (grant C14/22/083).

\bibliographystyle{apsrev4-1}
\bibliography{references}

\end{document}